\documentclass[a4paper, 11pt, amsmath, amssymb]{article}
\pdfoutput=1
\usepackage{jheppub}

\usepackage{epsfig}
\usepackage{dcolumn}
\usepackage{bm}
\usepackage{color}
\usepackage{hyperref}

\usepackage{amssymb}
\usepackage[figuresright]{rotating}
\usepackage{wrapfig}

\newcommand{\be}{\begin{equation}}
\newcommand{\ee}{\end{equation}}
\newcommand{\bea}{\begin{eqnarray}}
\newcommand{\eea}{\end{eqnarray}}

\title{Constraints on the Path-Length Dependence of Jet Quenching in
 Nuclear Collisions at RHIC and LHC}

\author[a]{Barbara Betz}
\author[b,c,d]{and Miklos Gyulassy}
\affiliation[a]{Institute for Theoretical Physics, Johann Wolfgang 
Goethe-University, 60438 Frankfurt am Main, Germany}
\affiliation[b]{Department of Physics, Columbia University, New York, 
10027, USA}
\affiliation[c]{Nuclear Science Division, Lawrence Berkeley National 
Laboratory, Berkeley, CA, USA}
\affiliation[d]{Institute for Particle and Nuclear Physics, Wigner RCP, 
HAS, 1121 Budapest, Hungary}

\emailAdd{betz@th.physik.uni-frankfurt.de}
\emailAdd{gyulassy@phys.columbia.edu}

\abstract{ 
Recent data on the high-$p_T$ pion nuclear modification factor, $R_{AA}(p_T)$, 
and its elliptic azimuthal asymmetry, $v_2(p_T)$, from  RHIC/BNL and LHC/CERN 
are analyzed in terms of a wide class of jet-energy loss models coupled to 
different (2+1)d transverse plus Bjorken expanding hydrodynamic fields. We test 
the consistency of each model by demanding a simultaneous account of the
azimuthal, the transverse momentum, and the centrality dependence of the data
at both 0.2 and 2.76 ATeV energies. We find a rather broad class of jet-energy 
independent energy-loss models $dE/dx= \kappa(T) x^z T^{2+z} \zeta_q$ that, 
when coupled to bulk constrained temperature fields $T(x,t)$, can account for 
the current data at the $\chi^2/{\rm d.o.f.}<2$ level with different temperature-dependent 
jet-medium couplings, $\kappa(T)$, and path-length dependence exponents 
$0\le z \le 2$. We extend previous studies by including a generic term, 
$0< \zeta_q < 2+q$, to test different scenarios of energy-loss fluctuations. 
While a previously proposed AdS/CFT jet-energy loss model with a 
temperature-independent jet-medium coupling as well as a near-$T_c$ dominated, 
pQCD-inspired energy-loss scenario are shown to be inconsistent with the LHC data, 
once the parameters are constrained by fitting to RHIC results, we find several 
new solutions with a temperature-dependent $\kappa(T)$. We conclude that the 
current level of statistical and systematic uncertainties of the measured data 
does not allow a constraint on the path-length exponent $z$ to a range narrower 
than $[0-2]$.
}


\begin{document}
\maketitle
\flushbottom

\section{Introduction} 
Jet-quenching observables \cite{glv,JET} have been proposed as tomographic 
probes of the density evolution of quark-gluon plasmas (QGPs) produced 
in high-energy nuclear collisions. It has been found that the nuclear 
modification pattern of jet distributions depends on a delicate complex 
interplay between the details of the jet-medium dynamics, 
$dE/dx = dE/dx[E(t),\vec{x}(t),T(t)]$, and the evolution of the bulk QGP 
collective temperature fields, $T(t)=T[\vec{x}(t),t]$. 

Below we investigate a wide variety of jet-energy loss models coupled to 
different QGP temperature fields constrained by bulk observables from 
state-of-the-art (viscous) (2+1)d hydrodynamic prescriptions 
\cite{Song:2008si, Shen:2010uy, Shen:2011eg, Luzum:2008cw, GVWH}. We compare
the obtained model results to recent data on the nuclear modification factor 
$R_{AA}(p_T,\phi,b,\sqrt{s})$ and the high-$p_T$ elliptic flow $v_2(p_T,\phi,b,\sqrt{s})$  
\cite{Adare:2012wg,Abelev:2012hxa,Abelev:2012di,CMS:2012aa,Chatrchyan:2012xq,ATLAS:2011ah}, 
investigating the transverse momentum $p_T$, the azimuthal $\phi$, 
the centrality $b$, and the collision energy $\sqrt{s}$ dependence of
the data measured at both the Relativistic Heavy Ion Collider (RHIC) and the 
Large Hadron Collider (LHC) with a special focus on the robustness of 
results for high-$p_T>7-10$~GeV hadron fragments from jets.

In particular, we study results based on a class of jet-energy loss models that 
can be parametrized as $dE/dx= \kappa(T) E^{a} x^z T^{c=2+z} \zeta_q$. 
The jet-energy dependence, the path-length dependence, and the temperature 
dependence are characterized by the exponents $(a,z,c)$. The above form 
allows for different assumptions of the distribution of the relative energy-loss 
fluctuations through a multiplicative factor $\zeta_q$, specifying a parameter 
$q$ as discussed in the text below, and rendering the $(a,z,c,q)$-prescription. 
This class of models includes perturbative QCD (pQCD) based models with exponents 
$(0,0,2,q)$ and $(0,1,3,q)$, conformal AdS holography models with non-linear path 
length $(0,2,4,q)$, and a phenomenological model assuming an enhancement of 
the jet-energy loss near $T_c\approx170$~MeV as in Ref.\ \cite{Liao:2008dk}, 
here referred to as the SLTc model with $(0,1,3,q)$. 

Since we find (see Tables \ref{tablesurvey1} and \ref{tablesurvey2} in the appendix) 
that all these models have problems to simultaneously account for the various data
measured both at RHIC and LHC, once the parameters are fixed at RHIC energies, we consider 
different deformations of those models by varying the assumed temperature
dependence of the jet-medium coupling, $\kappa(T)$. We confirm previous 
results pointing to the need to reduce of the jet-medium coupling from LHC to RHIC 
\cite{Betz:2012qq,WHDG11,Buzzatti:2012dy,Zakharov:2012fp,Lacey:2012bg,Pal:2012gf},
with a jet-medium coupling $\kappa_{\rm LHC}\approx 0.5\, \kappa_{\rm RHIC}$ that
negates most of the increase of the jet-energy loss as expected from the 
factor of $\sim 2$ increase of the QGP density at LHC relative to RHIC. 

However, even after that reduction is taken into account to describe the $R_{AA}$
at in the $p_T\sim 10$ GeV range, its elliptic azimuthal moment, $v_2$, is still 
found to be significantly underestimated by most models, especially at the LHC,
in line with various pQCD-based models (AMY, HT, ASW, Molnar, CUJET2.0) 
\cite{Adare:2012wg,Molnar:2013eqa,Xu14} that are about a factor of $\sim 2$ below
the measured data and might depend on the running of the coupling constant with both, the 
temperature and the scale $\alpha_{\rm eff}(Q,T)$ \cite{Xu14,Kaczmarek:2004gv}.
We therefore further explore deformations of the models that could help to resolve 
this ``high-$p_T$ $v_2$-problem''.   

We find that the pQCD-based models describing a vacuum running coupling with radiative 
energy-loss, $(0,1,3,q)$, require only a modest 10-15\% difference between the 
path-averaged coupling in- and out-of reaction plane. For AdS-like models, 
$(0,2,4,q)$, strong {\em non-conformal} temperature variations are required to 
bring those predictions closer to the combined RHIC and LHC data. Finally, a more 
radical deformation of the SLTc model with an exponential suppression of 
high-temperature jet-energy loss is reported that appears to be consistent 
within the present experimental and theoretical errors at RHIC and LHC.

The present work was motivated in part by a recent PHENIX study \cite{Adare:2012wg} 
suggesting the tentative conclusion that an AdS/CFT-motivated jet-energy loss 
prescription with $dE/dx\sim\kappa x^2T^4$ \cite{Bass:2008rv,Marquet:2009eq,Jia:2011pi} 
coupled to a particular hydrodynamic background \cite{Bass:2008rv} 
is more consistent with the observed azimuthal asymmetries than 
results based on pQCD radiative $(0,1,3,q)$ and elastic $(0,0,2,q)$
energy-loss models at RHIC energies used by AMY, HT, and ASW cited in Ref.\ 
\cite{Adare:2012wg}.   

Here, we test the consistency of those models, parametrized by $(a,z,c,q)$, with 
the observed $\sqrt{s}$ dependence between RHIC and LHC. As mentioned above, 
LHC jets probe the QGP phase of matter with up to an order of magnitude higher 
$p_T$-range under conditions where the QGP density $\propto T^3$ is more than 
doubled relative to RHIC. In addition, cross comparisons between RHIC and LHC 
are useful as the initial invariant jet-production distributions at midrapidity, 
$y=0$, denoted here as  $g_r(p_T)=dN_r^{jet}/dyd^2p_T$ for $r=q,g$ jets, changes 
by orders of magnitude from RHIC to LHC. Therefore, demanding a simultaneous 
description of RHIC and LHC data provides the most stringent test so far of the 
consistency and quantitative predictive power of proposed dynamical models of 
jet-energy loss and of the space-time evolution of the bulk QGP density produced 
in ultra-relativistic heavy-ion collisions.
 
\begin{figure}[tbh]
\hspace*{-1cm}
\includegraphics[width=6in]{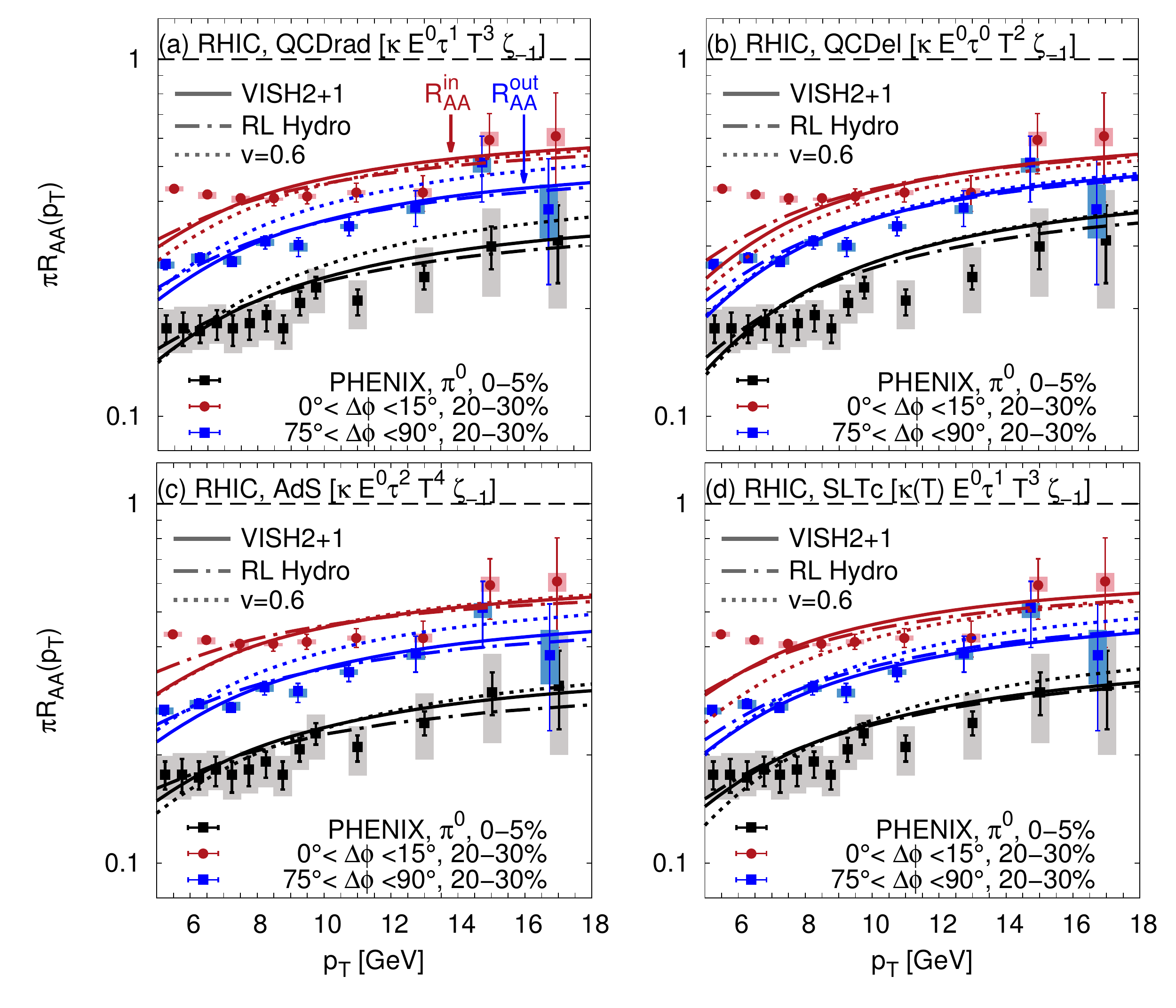}
\caption{Azimuthal jet tomography at RHIC. Panels (a-d) show PHENIX 200AGeV 
Au+Au data \protect{\cite{Adare:2012wg}} on the $\pi^0$ nuclear modification 
factors for most central 0-5\% collisions (black lines) as well as their in- and 
out-of-plane contributions for 20-30\% centralities (red and blue lines), compared 
to predictions based on $dE/dx=\kappa E^{a=0} x^z T^{c=2+z}\zeta_{-1}$
\protect{\cite{WHDG11, Betz:2012qq}} {\em without} energy loss fluctuations, 
i.e., $\zeta_{-1}=1$. Panel (a), labelled QCDrad $(z,c,q)=(1,3,-1)$, corresponds 
to a radiative pQCD-energy loss including running-coupling effects 
\protect{\cite{Betz:2012qq, Zakharov:2012fp, Buzzatti:2012dy, Xu14}}, panel 
(b), denoted QCDel with $(z,c,q)=(0,2,-1)$, describes an elastic jet-energy 
scenario \protect{\cite{WHDG11,DGLV}}, panel (c), marked as AdS with 
$(z,c,q)=(2,4,-1)$, simulates an idealized conformal falling string energy loss 
\protect{\cite{Gubser:2008as,Marquet:2009eq}}, and panel (d), indicated as SLTc 
with $(z,c,q)=(1,3,-1)$ and $\kappa(T_c) =3 \kappa(\infty)$, simulates a 
$T_c$-dominated energy-loss model proposed in Ref.\ \protect{\cite{Liao:2008dk}}. 
For each model, the quenching pattern is computed using three different bulk QGP 
fluid-temperature fields: ideal ($\eta/s=0$) VISH2+1 \protect{\cite{Song:2008si}} 
(solid), viscous ($\eta/s=0.08$) RL Hydro \protect{\cite{Luzum:2008cw}} 
(dashed-dotted), and a simple $v_\perp=0.6$ transverse blast wave model 
\protect{\cite{GVWH}} (dotted).}
\label{Fig1}
\end{figure}

Both, the magnitude and the azimuthal dependence of jet quenching in non-central 
collisions are conveniently studied via the nuclear modification factors
in- and out-of-plane $R_{AA}^{\rm in/out}=R_{AA}(1\pm 2v_2)$ 
\cite{Adare:2012wg,Bass:2008rv}, giving simultaneous access to both the nuclear 
modification factor and the high-$p_T$ elliptic flow. These observables are 
sensitive to all details of the jet energy, the path length, and the temperature 
dependence of jet-energy loss models (see, e.g.\ Refs.\ 
\cite{GVWH, Renk:2011aa,Renk:2011ia, Renk:2011gj,Chen:2011vt, Molnar:2013eqa}).
In particular, they depend on the details of the QGP transverse as well as 
longitudinal expansion \cite{Song:2008si,Shen:2010uy,Shen:2011eg,Luzum:2008cw,Niemi:2008ta},
as emphasized by Renk \cite{Renk:2011aa} and Molnar \cite{Molnar:2013eqa}. 

The jet-medium coupling $\kappa$ used is constrained for each model by fitting
to a single reference point at $p_T=7.5$~GeV in most central Au+Au at 
$\sqrt{s}=200$~AGeV RHIC energies with the value of $R_{AA}(p_T)=0.2$, 
as in Refs.\ \cite{Bass:2008rv,Fries:2010jd,Betz:2012qq}. However, results are 
found to be qualitatively insensitive to the particular choice of the 
$p_T$-reference point. 

We generalize our previous work \cite{Betz:2012qq} by including more realistic 
energy-loss fluctuations to our model and discussing the implications of the
high-$p_T$ $v_2$-problem on the jet-medium coupling.

\begin{figure}[t]
\hspace*{-2ex}
\includegraphics[width=6.1in]{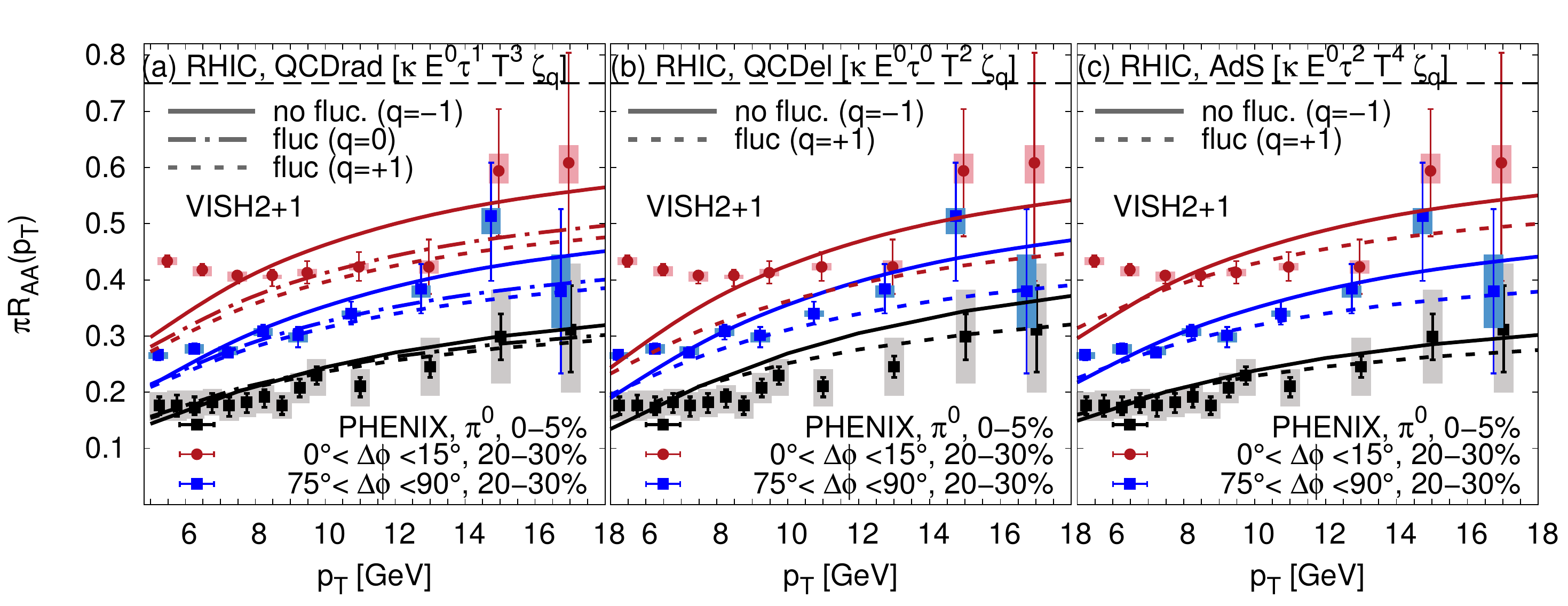}
\caption{Azimuthal jet tomography at RHIC, comparing predictions based on 
$dE/dx=\kappa E^{a=0} x^z T^{c=2+z}\zeta_{q}$ 
\protect{\cite{WHDG11, Betz:2012qq}} with and without energy fluctuations 
to PHENIX 200AGeV Au+Au data \protect{\cite{Adare:2012wg}} on $\pi^0$ 
nuclear modification factors for most central 0-5\% collisions and their 
in- and out-of-plane contributions for 20-30\% centralities. Scenarios 
(a)-(c) are the same as in Fig.\ \ref{Fig1}, computed for the ideal VISH2+1 
bulk QGP fluid field \protect{\cite{Song:2008si}}.}
\label{Fig2}
\end{figure}

\begin{figure}[b]
\hspace*{-1cm}
\begin{center}
\includegraphics[width=3in]{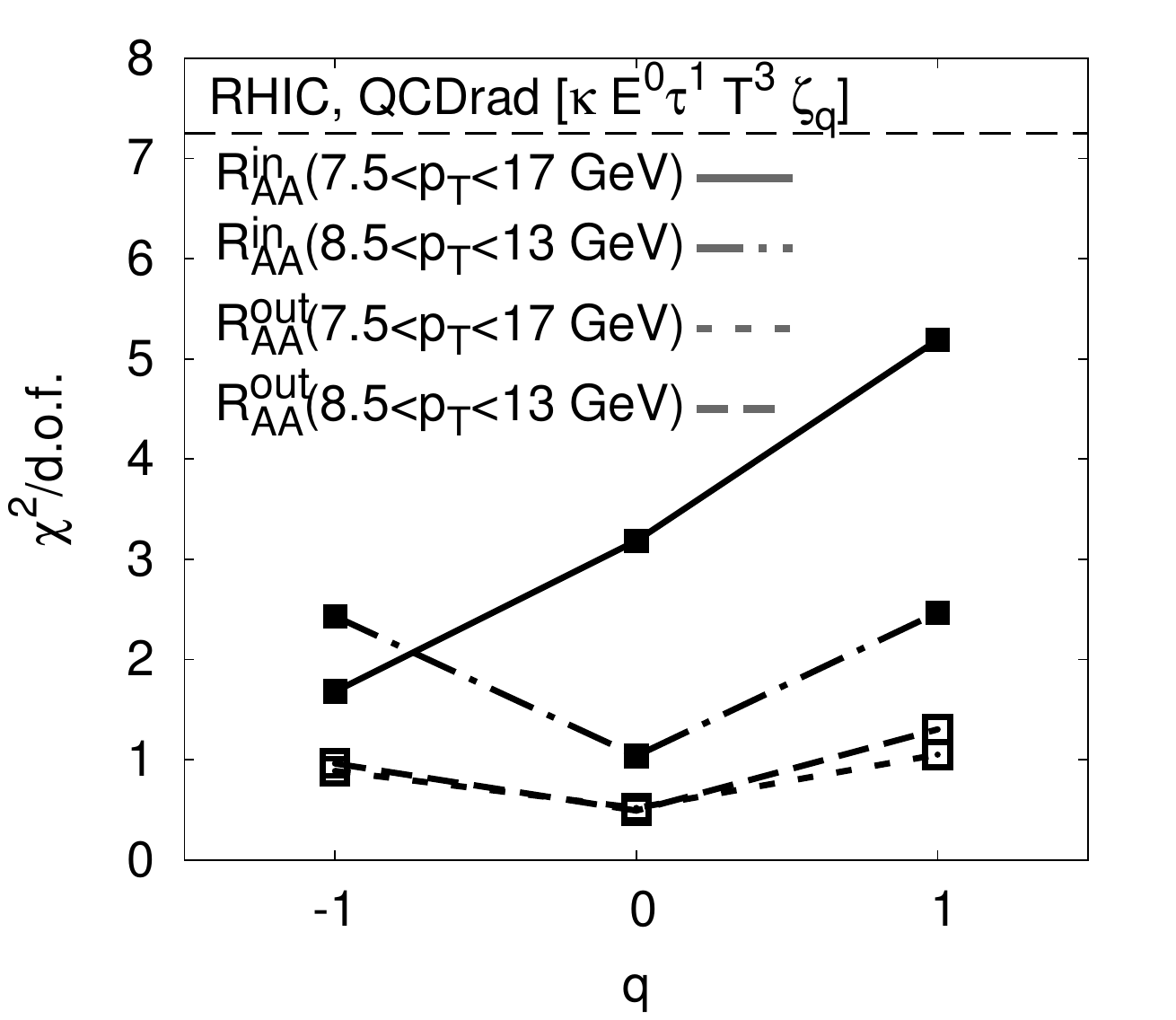}
\caption{The $\chi^2$/d.o.f.\ for the QCDrad scenario shown in Fig.\ \ref{Fig2} 
(a) as a function of the three energy-loss fluctuation distributions 
$q=[-1,0,1]$ considering VISH2+1 background fields \protect{\cite{Song:2008si}}. 
The different $p_T$-cuts demonstrate the sensitivity of a $\chi^2$-test on 
the $p_T$-range considered.}
\label{Fig3}
\end{center}
\end{figure}

\section{The $(z,c,q)$-Classification of $dE/dx$-Models} 

In order to interpolate between the different jet-energy loss prescriptions,
we utilize a convenient parametric model introduced in Refs.\ \cite{WHDG11,Betz:2012qq} 
that originally characterized the jet-energy loss by three exponents $(a,z,c)$
controlling the jet energy $a$, path length $z$, and thermal-field dependence $c$.
Here, we allow for additional energy-loss fluctuations $\zeta_q$, discussed below, 
and the possibility that the jet-medium coupling, $\kappa(T)$, may depend 
non-monotonically on the local temperature field,
\begin{eqnarray}
\hspace*{-3ex}
\frac{dE}{dx}=\frac{dP}{d\tau}(\vec{x}_0,\phi,\tau)= 
-\kappa(T)  P^a(\tau) \, \tau^{z} \, T^c \, \zeta_q
\;,
\label{Eq1}
\end{eqnarray}
where $\kappa(T)=C_r\kappa^\prime(T)$ and 
$T=T[\vec{x}(\tau)=\vec{x}_0+ (\tau-\tau_0) \hat{n}(\phi),\tau]$
describes the local temperature along the jet path at time $\tau$ for a jet
initially produced time $\tau_0$. The jets are distributed according to
a transverse initial profile specified by the bulk QGP flow fields given by 
three variants of transverse plus Bjorken $(2+1)$d expansion: (1) VISH2+1 
\cite{Song:2008si,Shen:2011eg}, (2) viscous RL hydro \cite{Luzum:2008cw}, 
and (3) a $v_\perp=0.6$ blast wave flow \cite{GVWH} assuming radial 
dilation of the initial transverse profile:
$\rho(x,y,\tau)=\rho_0[x/r(\tau),y/r(\tau)][\tau_0/\tau r^2(\tau)]$
with $r(\tau)=(1+v_\perp^2\tau^2/R^2)^{1/2}$.
Here, $R$ denotes the initial root mean square radius. For dimensionless 
couplings $\kappa$, $c=2+z-a$.  In Eq.\ (\ref{Eq1}), $C_r=1(\frac{9}{4})$ 
describes quark (gluon) jets. For jets of type $r=q,g$ produced with an invariant
transverse momentum distribution $g_r(P_0)$ taken from Refs.\ 
\cite{WHDG11,Betz:2012qq}, the nuclear modification factor is given by
\begin{eqnarray}
R_{AA}^r(P_f,\phi)=\frac{\langle g_r[P_0^r(P_f,\phi)]\rangle}{
g_r(P_f)} \frac{dP^2_0}{dP^2_f}
\; .
\end{eqnarray}
The ensemble average is taken over initial jet-production points and
initial $\sqrt{s}$-dependent initial jet energies $P_0$, as well as 
parameters controlling the energy loss, geometry, and temperature 
background fields.

\begin{figure*}[t]
\includegraphics[width=6in]{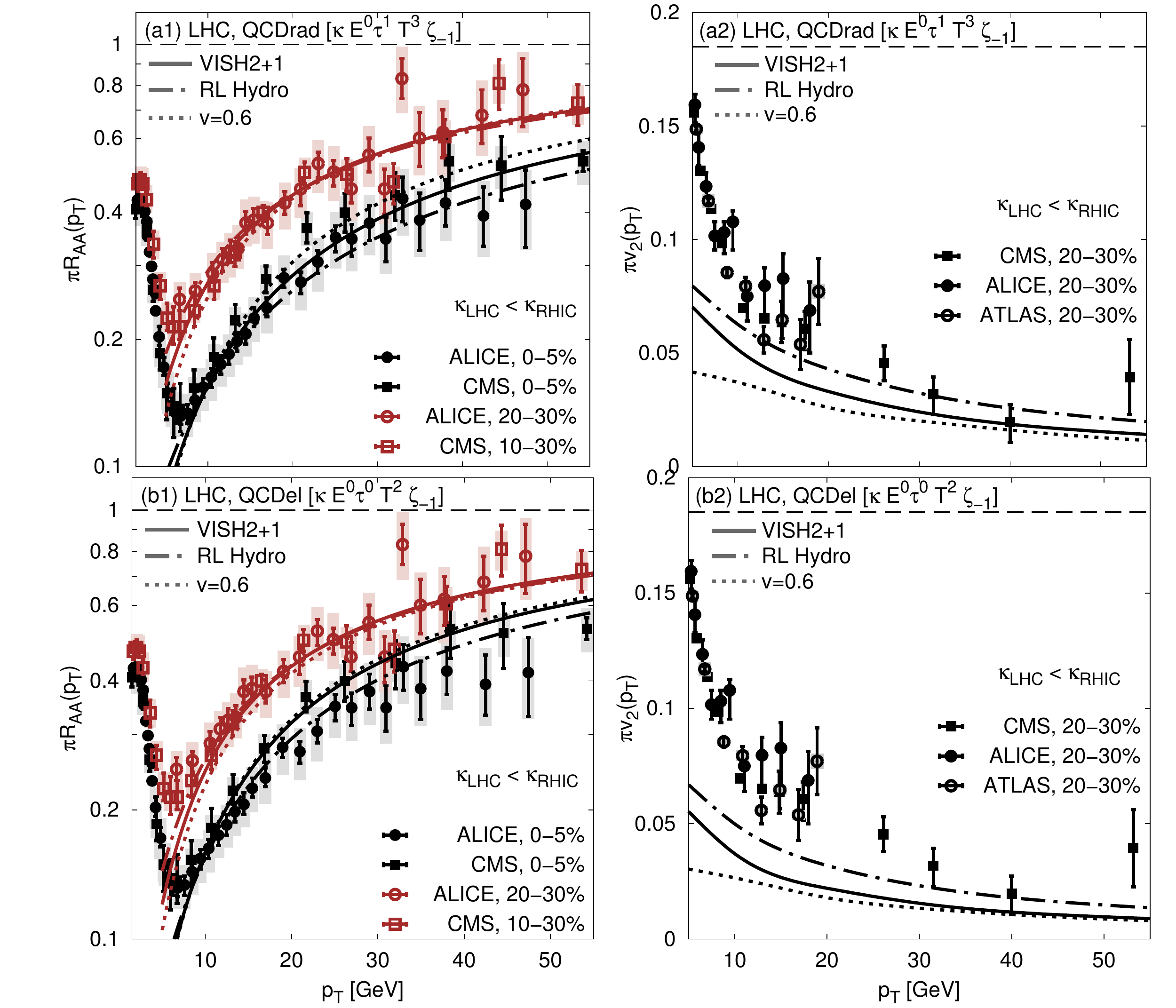}
\caption{Azimuthal jet tomography at the LHC 
\cite{Abelev:2012hxa,Abelev:2012di,CMS:2012aa,Chatrchyan:2012xq,ATLAS:2011ah}. 
Panels (a1) and (b1) show the measured data for the pion nuclear modification 
factor $R_{AA}$ from ALICE \protect{\cite{Abelev:2012hxa}} and CMS 
\protect{\cite{CMS:2012aa}} for most central and more peripheral collisions, 
while panels (a2) and (b2) depict the high-$p_T$ elliptic flow as extracted from 
ALICE \protect{\cite{Abelev:2012di}}, CMS \protect{\cite{Chatrchyan:2012xq}}, 
and ATLAS \protect{\cite{ATLAS:2011ah}}. The model calculations are done 
{\em without} energy-loss fluctuations ($\zeta_{-1}=1$) to mimic a radiative 
QCD (QCDrad, upper panel) and an elastic QCD (QCDel, lower panel) energy loss 
using bulk QGP flow fields at LHC energies from viscous $\eta/s=0.08$ VISH2+1 
\protect{\cite{Shen:2011eg}} (solid), viscous $\eta/s=0.08$ RL Hydro 
\protect{\cite{Luzum:2008cw}} (dashed-dotted), and the $v_\perp=0.6$ blast 
wave model \protect{\cite{GVWH}} (dotted). The jet-medium coupling 
$\kappa_{LHC}$ is reduced relative to RHIC to simulate the running QCD coupling 
\protect{\cite{Buzzatti:2012dy,Zakharov:2012fp, Kaczmarek:2004gv}}.}
\label{Fig4}
\end{figure*}

\begin{figure*}[t]
\includegraphics[width=6in]{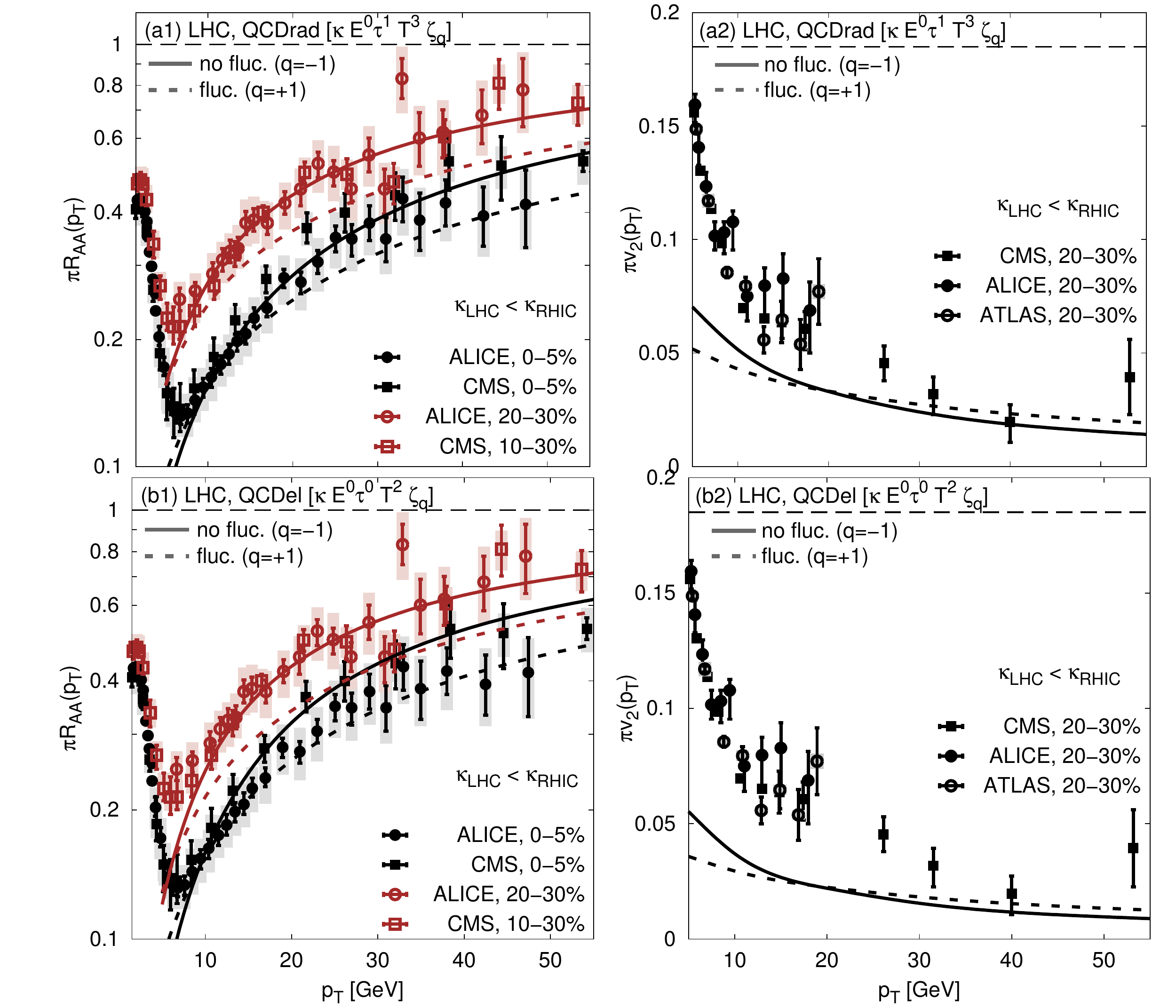}
\caption{Azimuthal jet tomography at the LHC 
\cite{Abelev:2012hxa,Abelev:2012di,CMS:2012aa,Chatrchyan:2012xq,ATLAS:2011ah}
as in Fig.\ \ref{Fig4}. The model calculations are done for a reduced $\kappa$ 
value as compared to RHIC, comparing a fluctuating energy-loss scenario (dashed) 
to the non-fluctuating case (solid) using bulk QGP flow fields at LHC energies 
from viscous $\eta/s=0.08$ VISH2+1 \protect{\cite{Shen:2011eg}}.}
\label{Fig5}
\end{figure*}

\begin{figure}[t]
\begin{center}
\includegraphics[width=6in]{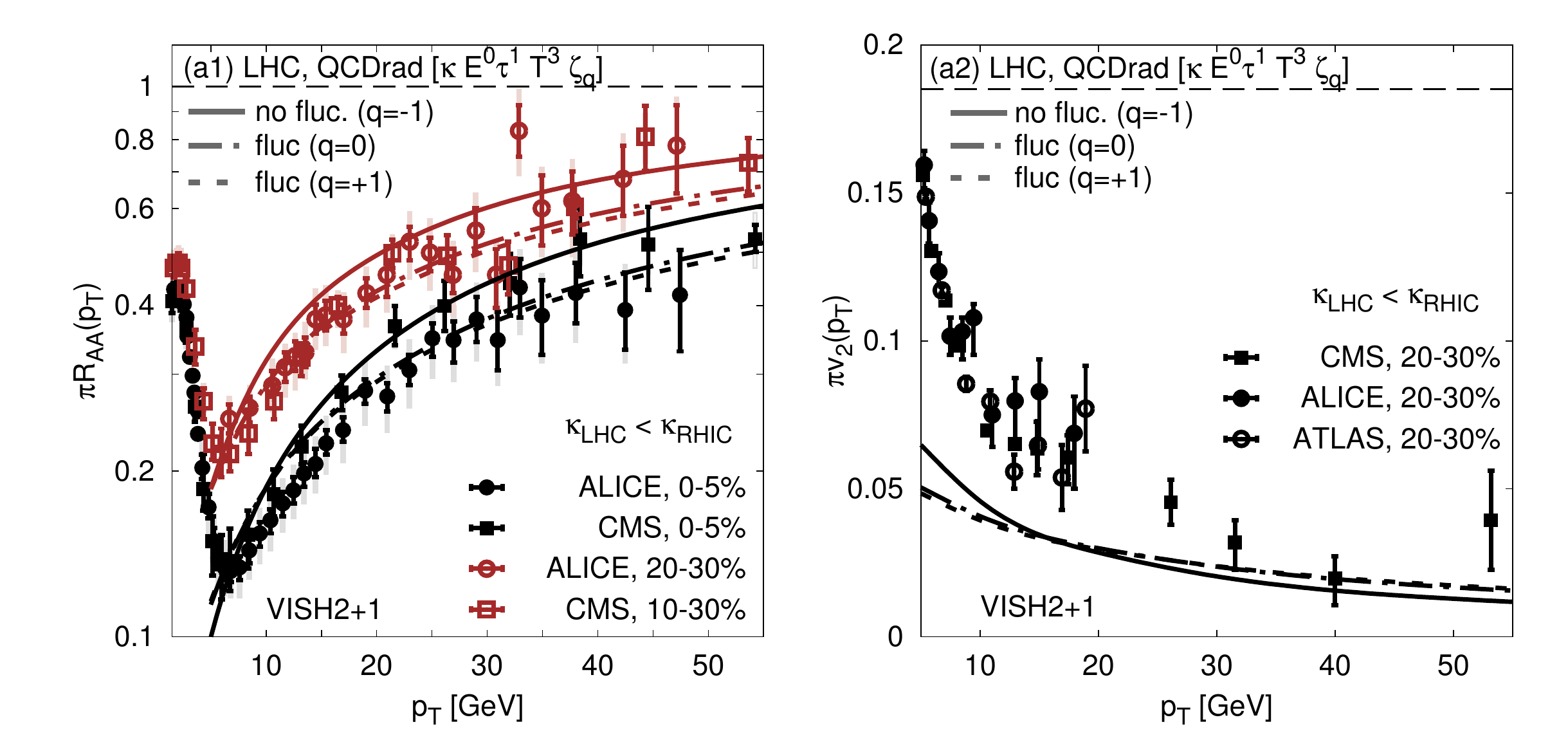}
\caption{Azimuthal jet tomography at the LHC 
\cite{Abelev:2012hxa,Abelev:2012di,CMS:2012aa,Chatrchyan:2012xq,ATLAS:2011ah}
as in Fig.\ \ref{Fig5} (a1) and (a2) considering both fluctuating 
and non-fluctuating jet-energy loss scenarios ($q=-1,0,1$) but assuming 
that $R_{AA}(p_T=10\,{\rm GeV})=0.186$ instead of 
$R_{AA}(p_T=10\,{\rm GeV})=0.155$ as in Fig.\ \ref{Fig5}.}
\label{Fig6}
\end{center}
\end{figure}

\begin{figure}[b]
\begin{center}
\includegraphics[width=3in]{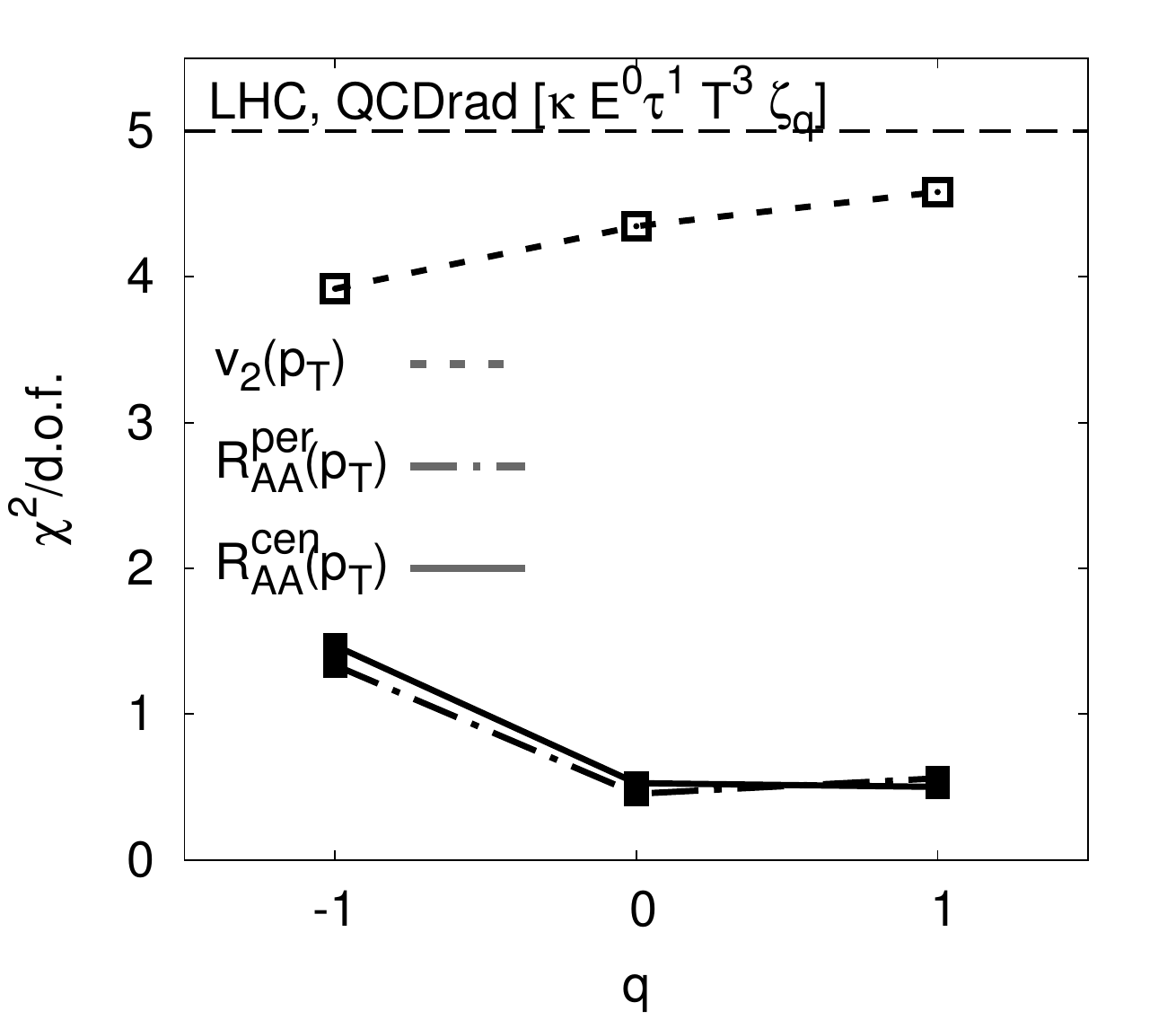}
\caption{The $\chi^2$/d.o.f.\ for the QCDrad scenario considering VISH2+1 
\protect{\cite{Shen:2011eg}} background fields shown in Fig.\ \ref{Fig6} 
as a function of the three energy-loss fluctuation distributions 
$\zeta_{q}=[-1,0,1]$. For the ALICE data, a $p_T$-range of $10<p_T<48$~GeV is 
considered for the pion nuclear modification factor and $10<p_T<20$~GeV for the 
high-$p_T$ elliptic flow, while for the CMS data the $p_T$-cut of $10<p_T<54$~GeV 
holds both for the nuclear modification factor and the high-$p_T$ elliptic flow.}
\label{Fig7}
\end{center}
\end{figure}

For a given member of the jet ensemble, the average initial jet energy $P_0^r$, 
is related to the final quenched jet energy (prior to hadronization) $P_f$
via a path integral that we here assume to be an Eikonal straight line in 
azimuthal direction $\phi$. For a particular jet flavor $r$, the average initial 
jet energy is
\begin{eqnarray} 
\hspace*{-1ex} 
P_0^r(P_f,\phi)=\left[P_f^{1-a}+ \zeta_q \int_{\tau_0}^{\tau_f}
K_r(T) \tau^z T^{c}[\vec{x}_\perp(\tau),\tau]d\tau
\right]^\frac{1}{1-a}\hspace*{-2ex} \; ,
\label{Eq2}
\end{eqnarray} 
with the effective jet-medium coupling $K_r(T)=(1-a)C_r \kappa(T)$. 
Eq.\ (\ref{Eq2}) illustrates the competing effects due to the intrinsic 
$dE/dx\propto E^ax^zT^c$ energy-loss details, the impact of local 
hydrodynamic temperature fields, and a possible non-monotonic jet-medium coupling 
$\kappa(T[\vec{x}(t),t])$ along its path. We checked numerically that local 
transverse flow-field effects introduced in Ref.\ \cite{Baier:2006pt} 
do not significantly influence the results based on Eq.\ (\ref{Eq2}).

With Eq.\ (\ref{Eq2}) we have further generalized the class of $(a,z,c)$-models 
to $(a,z,c,q)$-models that include the possibility of skewed jet-energy loss 
fluctuations about its path-averaged mean using a scaling factor 
$0<\zeta_q< q+2$ and being distributed according to
\be
f_q(\zeta_q)= \frac{(1 + q)}{(q+2)^{1+q}} (q + 2- \zeta_q)^q 
\label{fq}
\ee
with a root-mean square of $\langle( \zeta_q-1)^2 \rangle=(q+1)/(q+2)$.
This class of skewed distributions is controlled by a parameter $q>-1$
with unit mean, $\langle \zeta_q \rangle=1$. It conveniently interpolates 
between non-fluctuating ($q=-1$, $\zeta_{-1}=1$), uniform Dirac
$\delta(1-\zeta_{-1})=\lim_{q\rightarrow -1^+}f_q(\zeta_{-1})$
distributions between $0<\zeta_{-1} <1$, and increasingly skewed distributions
towards small $\zeta_q < 1$ for $q>-1$ similar to pQCD based models, 
see e.g., Refs.\ \cite{glv,WHDG11, DGLV}. Note that current non-perturbative 
AdS and the originally proposed SLTc models \cite{Liao:2008dk} do not include 
fluctuations of the jet-energy loss about its path average and thus correspond 
to the $q=-1, \zeta_{-1}=1$ limit of Eq.\ (\ref{fq}). 

Including the energy-loss fluctuations specified by Eqs.\ (\ref{Eq2}) and (\ref{fq})
into Eq.\ (\ref{Eq1}) thus conveniently classifies jet-medium models labelled by
$(a,z,c,q)$ to differentiate between a much broader class of jet-energy loss 
models than reported in Ref.\ \cite{Betz:2012qq}. By varying these four model space 
parameters we aim to quantify the exponents and to identify which combinations 
of jet-energy loss and bulk QGP evolution models can be ruled out by the current 
RHIC and LHC data.

We limit the study to the special cases $a=0$, $z=[0,1,2]$, $c=2+z$, and 
$q=[-1,0,1]$, and hence the ($z,c,q$)-model. Our restriction to $a=0$ and thus 
jet-energy independent energy-loss models is motivated by earlier results 
reported in Ref.\ \cite{Betz:2012qq} showing that the slope of 
$R_{\rm PbPb}(10<p_T<40\,{\rm GeV}, 0-10\%\,{\rm centrality}, \sqrt{s}=2.76\,{\rm ATeV})$
strongly disfavors models with $a>1/3$. 

In particular, we investigate a pQCD-like radiative energy loss QCDrad with 
$dE/dx=\kappa E^0\tau^1T^3\zeta_q$, a pQCD-like elastic energy loss QCDel with 
$dE/dx=\kappa E^0\tau^0T^2\zeta_q$, an AdS/CFT-inspired scenario with 
$dE/dx=\kappa E^0\tau^2T^4\zeta_q$, and a SLTc model with 
$dE/dx=\kappa(T) E^0\tau^1T^3\zeta_q$ that has a coupling
constant with a constant, non-negligible value for large temperatures that is 
enhanced around $T\sim T_c\approx 170$~MeV \cite{Liao:2008dk}. Besides that, 
we also study two models with a jet-medium coupling that either depends on the 
azimuth \cite{Xu14} or drops exponentially for large temperates. We consider a 
weaker $\kappa_{\rm LHC}<\kappa_{\rm RHIC}$ coupling at LHC energies for the 
pQCD-like jet-energy loss prescriptions 
\cite{Betz:2012qq,Xu14,WHDG11,Buzzatti:2012dy,Zakharov:2012fp}.

\section{RHIC and LHC Results}
Fig.\ \ref{Fig1} shows the most central (black lines) and more peripheral pion
nuclear modification factors in- and out-of-plane (red and blue lines) for the
first four model scenarios introduced above in panels (a) to (d) not considering energy-loss
fluctuations, i.e.\ $\zeta_{-1}=1$, for the three $(2+1)$d flow background 
fields of ideal VISH2+1 (solid) \cite{Song:2008si}, viscous RL 
Hydro (dashed-dotted) \cite{Luzum:2008cw}, and the $v_\perp=0.6$ blast wave 
flow \cite{GVWH}. Please note that in Refs.\ \cite{Buzzatti:2012dy} 
the opacity integral of Eq.\ (\ref{Eq2}) was evaluated taking only Bjorken 
expansion with a $v_\perp=0$ into account. 

The most striking result in Fig.\ \ref{Fig1} is that in contrast to the (AMY,
HT, and ASW) pQCD models \cite{Bass:2008rv} shown in Ref.\ \cite{Adare:2012wg}, 
all models combined with either ideal VISH2+1 or viscous RL Hydro transverse 
flow fields agree within present errors with the measured RHIC data in the 
high-$p_T> 7$ GeV region. Only the QCDel model seems to be disfavored as
compared to the other scenarios. We checked (not shown) that the results of viscous 
VISH2+1 \cite{Shen:2010uy} background fields vary by less than 5\%. 
However, the $v_\perp=0.6$ transverse blast wave background leads, as in 
Ref.\ \cite{GVWH} with $v_\perp =0$, to an in/out asymmetry with a factor of 
$\sim$two below the recent PHENIX data \cite{Adare:2012wg}. Ref.\ 
\cite{Molnar:2013eqa} also reports that the GLV energy-loss \cite{glv} 
evaluated in the MPC parton cascade background under-predicts the high-$p_T$ 
elliptic asymmetry observed at RHIC. This result \cite{Molnar:2013eqa} was
another major motivation for the present work, as well as the detailed investigation
provided by CUJET2.0 \cite{Xu14}.

The differences between the models shown in Fig.\ \ref{Fig1} and the results reported
in Refs.\ \cite{Adare:2012wg} are due to various combined effects of the
jet-energy loss and the bulk QGP flow. The flow fields \cite{Bass:2008rv} considered 
for the study in Ref.\ \cite{Adare:2012wg} were computed with an ideal 
(non-dissipative) hydrodynamic code assuming a Bag model first order-phase 
transition with a speed of sound vanishing over a wide energy-density range. 
Here, however, the VISH2+1 results used in Fig.\ \ref{Fig1} utilizes 
a smoothed (SM-EOS Q) equation of state (EoS) and the viscous RL Hydro employs 
a realistic continuous crossover transition EoS. 

Besides that, the results shown in Fig.\ \ref{Fig1} do not include jet-energy
loss fluctuations as deduced in Eq.\ (\ref{Eq2}) that are intrinsic to the models
reported in Refs.\ \cite{Xu14,Molnar:2013eqa}. This effect is, however, 
included in Fig.\ \ref{Fig2} for the three scenarios of QCDrad, QCDel, and AdS, 
comparing the non-fluctuating case of $q=-1$ to two different energy-loss fluctuations
given by $q=0,1$. For reasons of clarity and to better compare to results
of CUJET2.0 \cite{Xu14}, we restricted this comparison to the VISH2+1 
background fields \cite{Shen:2011eg} and omitted the SLTc model as the effects 
of a fluctuating jet-energy loss for a model with $dE/dx\sim E^0\tau^1T^3\zeta_q$ 
is already illustrated by the QCDrad scenario shown in Fig.\ \ref{Fig2}(a). 
Please note that the detailed study of Ref.\ \cite{Xu14} demonstrated that the GLV 
energy-loss considered in both Refs.\ \cite{Xu14,Molnar:2013eqa} may account for 
the $v_2$ data at RHIC for an $\alpha_{max}=0.26-0.28$. The differences in 
the results of Ref.\ \cite{Xu14} and \cite{Molnar:2013eqa} can only be 
explained in the details of the (running) coupling constant as well as the
background medium considered.

\begin{figure*}[b]
\begin{center}
\hspace*{-5ex}
\includegraphics[width=6in]{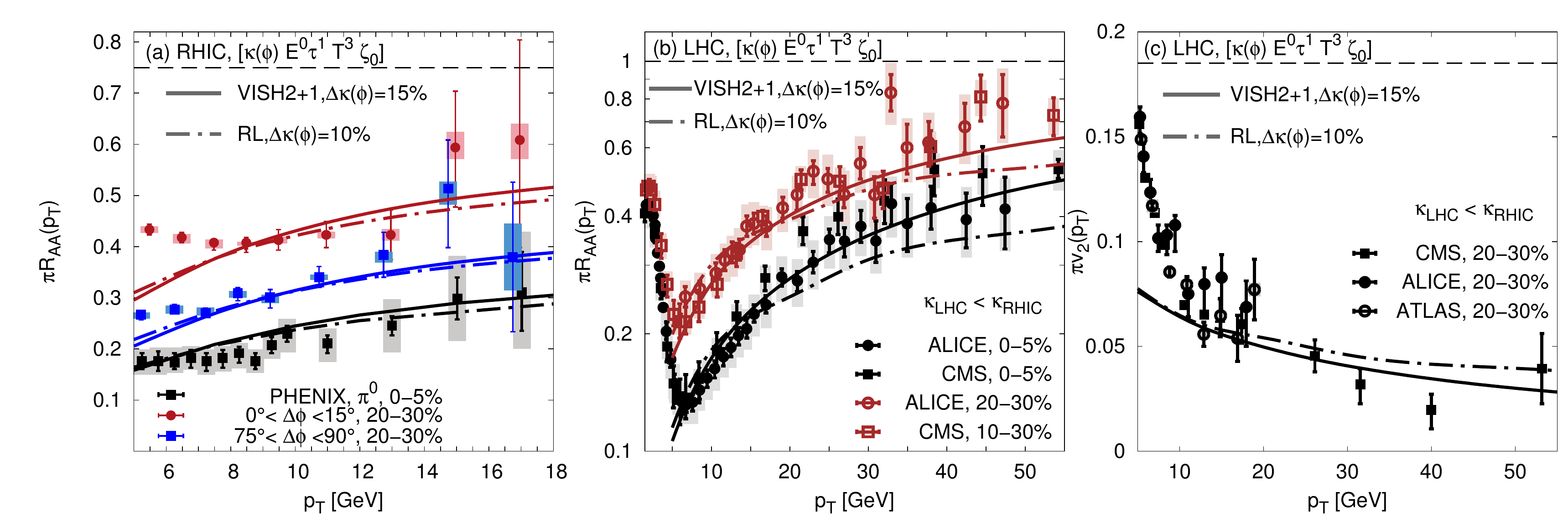}
\caption{Azimuthal jet tomography at RHIC and LHC assuming jet-energy loss 
fluctuations and a moderate azimuthal dependence of the jet-medium coupling of 
$\Delta\kappa(\phi)=15\%$ for the bulk QGP flow fields from VISH2+1 
\protect{\cite{Shen:2011eg}} (solid) and $\Delta\kappa(\phi)=10\%$ for the 
RL fields \protect{\cite{Luzum:2008cw}} (dashed-dotted). Panel (a) shows 
the nuclear modification factor for most central collisions as well as their 
in-and out-of-plane contributions at RHIC, panel (b) depicts the $R_{AA}(p_T)$ 
at LHC, and panel (c) describes the high-$p_T$ elliptic flow at LHC energies. 
At LHC energies, the jet-medium coupling is reduced as compared to RHIC energies 
to account for the QCD running-coupling effect.}
\label{Fig8}
\end{center}
\end{figure*}

\begin{figure*}[t]
\begin{center}
\includegraphics[width=6in]{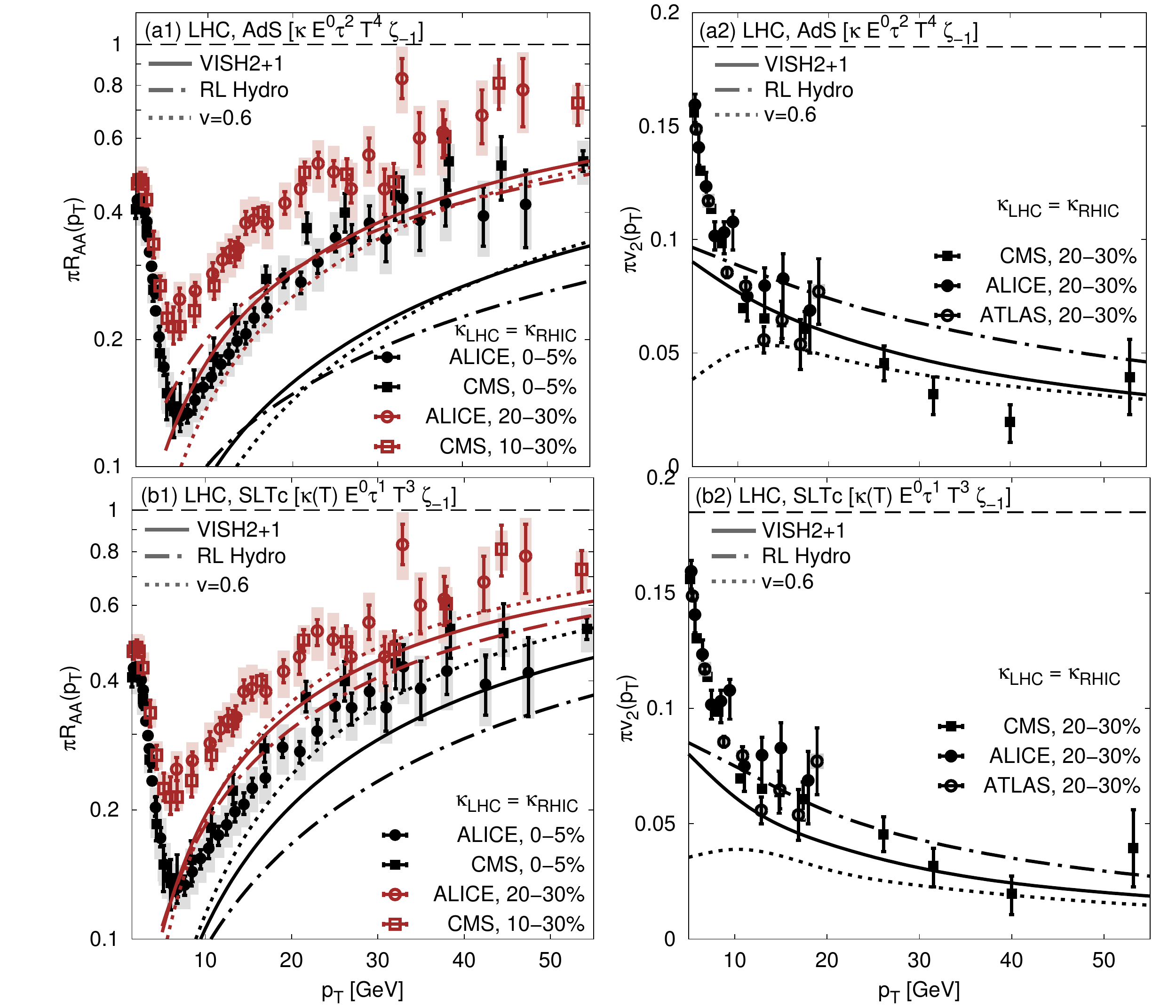}
\caption{Azimuthal jet tomography at the LHC 
\cite{Abelev:2012hxa,Abelev:2012di,CMS:2012aa,Chatrchyan:2012xq,ATLAS:2011ah}. 
Panels (a1) and (b1) show the measured data for the pion nuclear modification 
factor $R_{AA}$ from ALICE \protect{\cite{Abelev:2012hxa}} and CMS 
\protect{\cite{CMS:2012aa}} for most central and more peripheral collisions, 
while panels (a2) and (b2) depict the high-$p_T$ elliptic flow as extracted from
ALICE \protect{\cite{Abelev:2012di}}, CMS \protect{\cite{Chatrchyan:2012xq}},
and ATLAS \protect{\cite{ATLAS:2011ah}}. The model calculations are done to
mimic a {\em conformal} AdS \protect{\cite{Marquet:2009eq,Gubser:2008as}} 
(upper panel) and a $T_c$-dominated SLTc model \protect{\cite{Liao:2008dk}} 
(lower panel) {\em without} energy-loss fluctuations, using bulk QGP flow fields 
at LHC energies from viscous $\eta/s=0.08$ VISH2+1 \protect{\cite{Shen:2011eg}} 
(solid), viscous $\eta/s=0.08$ RL Hydro \protect{\cite{Luzum:2008cw}} 
(dashed-dotted), and the $v_\perp=0.6$ blast wave model \protect{\cite{GVWH}} 
(dotted). Here, the {\em same} jet-medium coupling $\kappa_{LHC}$ is taken at 
the LHC as fixed at RHIC, see text.}
\label{Fig9}
\end{center}
\end{figure*}

\begin{figure*}[t]
\begin{center}
\includegraphics[width=6in]{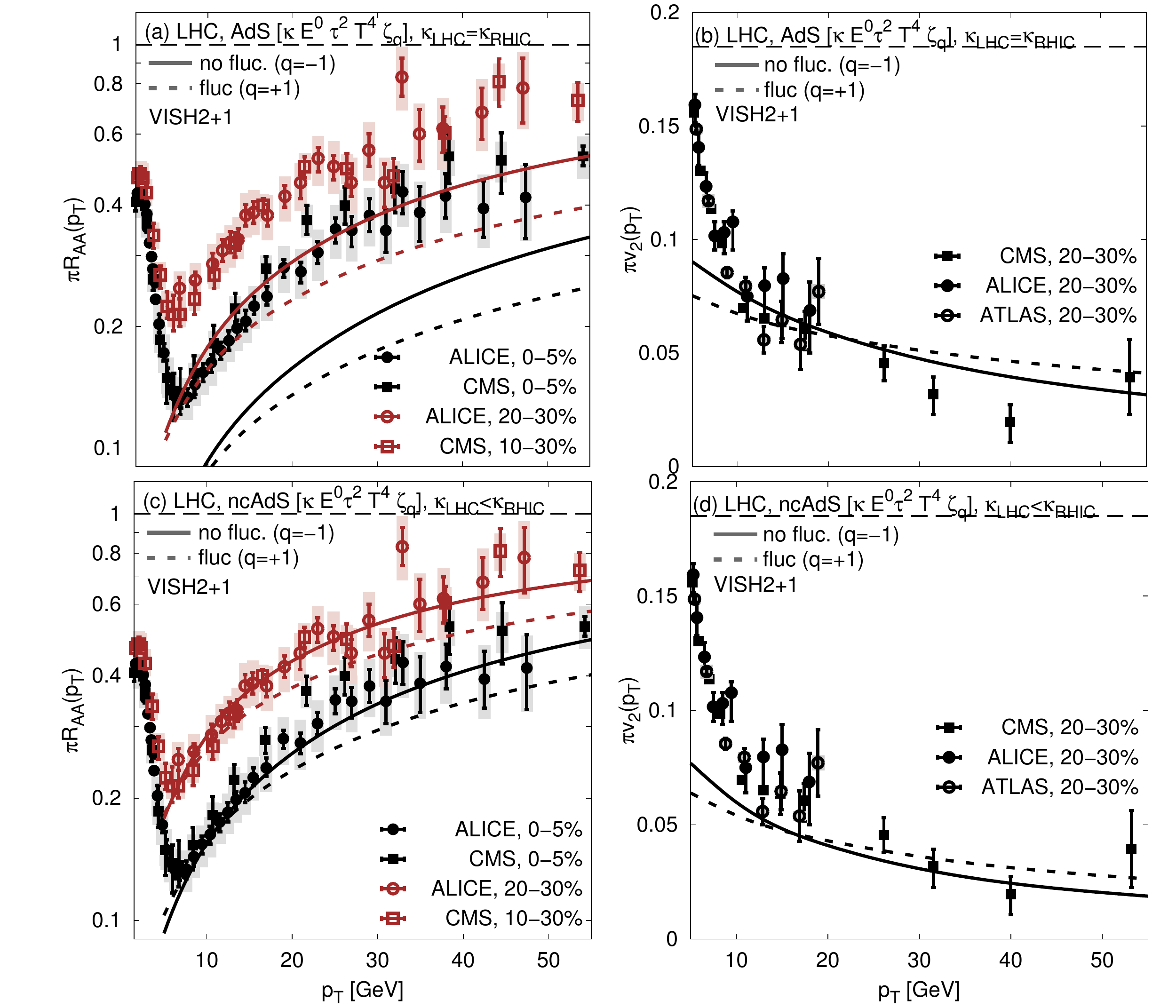}
\caption{Azimuthal jet tomography at the LHC 
\cite{Abelev:2012hxa,Abelev:2012di,CMS:2012aa,Chatrchyan:2012xq,ATLAS:2011ah}. 
Panels (a) and (c) show the measured data for the pion nuclear modification 
factor $R_{AA}$ from ALICE \protect{\cite{Abelev:2012hxa}} and CMS 
\protect{\cite{CMS:2012aa}} for most central and more peripheral collisions, 
while panels (b) and (d) depict the high-$p_T$ elliptic flow as extracted from
ALICE \protect{\cite{Abelev:2012di}}, CMS \protect{\cite{Chatrchyan:2012xq}},
and ATLAS \protect{\cite{ATLAS:2011ah}}. The model calculations are done for 
the $dE/dx=\kappa E^0\tau^2T^4\zeta_q$ scenario, comparing a fluctuating 
energy loss ($\zeta_{+1}$, dashed) to the non-fluctuating scenario ($\zeta_{-1}$, 
solid) using bulk QGP flow fields at LHC energies from viscous $\eta/s=0.08$ 
VISH2+1 \protect{\cite{Shen:2011eg}}. In the upper panel, the {\em same} 
$\kappa$ is taken as at RHIC, while in the lower panel, a {\em reduced} 
jet-medium coupling is assumed.}
\label{Fig10}
\end{center}
\end{figure*}

Including jet-energy loss fluctuations, both the QCDrad and the AdS model
depicted in Fig.\ \ref{Fig2} describe the measured data, while the QCDel 
scenario is again disfavored. Fig.\ \ref{Fig3} clearly demonstrates that the 
$q=0$ case reproduces the nuclear modification factor in- and out-of plane with 
a $\chi^2/{\rm d.o.f.}\leq 1.5$ if a $p_T$-range of $8.5<p_T<13$~GeV is considered. For a
 wider $p_T$-range, even the non-fluctuating $q=-1$ energy loss leads to a 
decent description with a $\chi^2/{\rm d.o.f.}< 2$. Please note that the applicability of 
our model is limited below $p_T=7.5$~GeV.

Thus, comparing the various models at RHIC energies only allows the conclusion 
that the QCDrad scenario with and without jet-energy loss fluctuations
as well as the AdS scenario and the SLTc model are possible candidates 
to describe the nuclear modification factor and the high-$p_T$ elliptic flow
while the QCDel scenario seems to be disfavored. 

Given this result as well as the difficulty of untangling the effect of the 
jet-energy loss and QGP flow fields at one particular collision energy 
$\sqrt{s}$ lead us to consider the higher discriminating power afforded by 
exploiting the dependence of the $R_{AA}$ and the high-$p_T$ elliptic flow on
the collision energy in the range of $0.2-2.76$~ATeV.

\begin{figure*}[t]
\begin{center}
\includegraphics[width=3in]{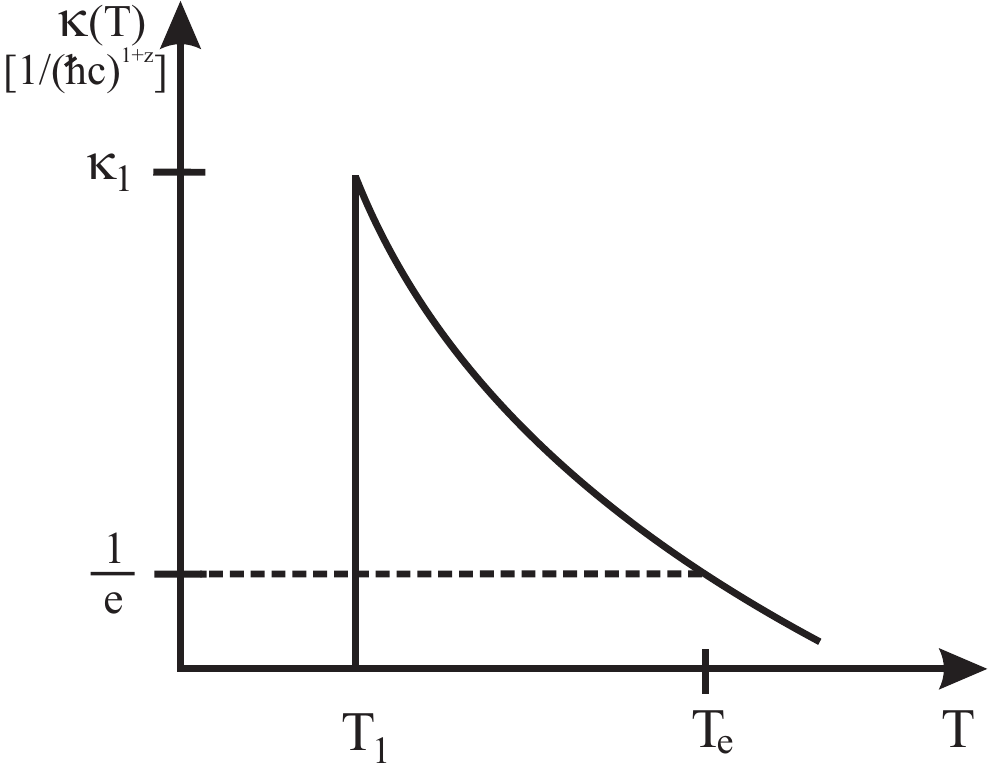}
\caption{Sketch of an exponential temperature-dependent jet-medium coupling
$\kappa(T)$, in units of $1/(\hbar c)^{(1+z)}$, as given by Eq.\ (\ref{k_exp}) 
assuming that the coupling is zero below a temperature $T_1$, peaks at $T_1$ 
with a value of $\kappa_1$, and falls off to a value of $1/e$ at a 
temperature $T_e$.}
\label{Fig11}
\end{center}
\end{figure*}

\begin{figure*}[b]
\begin{center}
\includegraphics[width=6in]{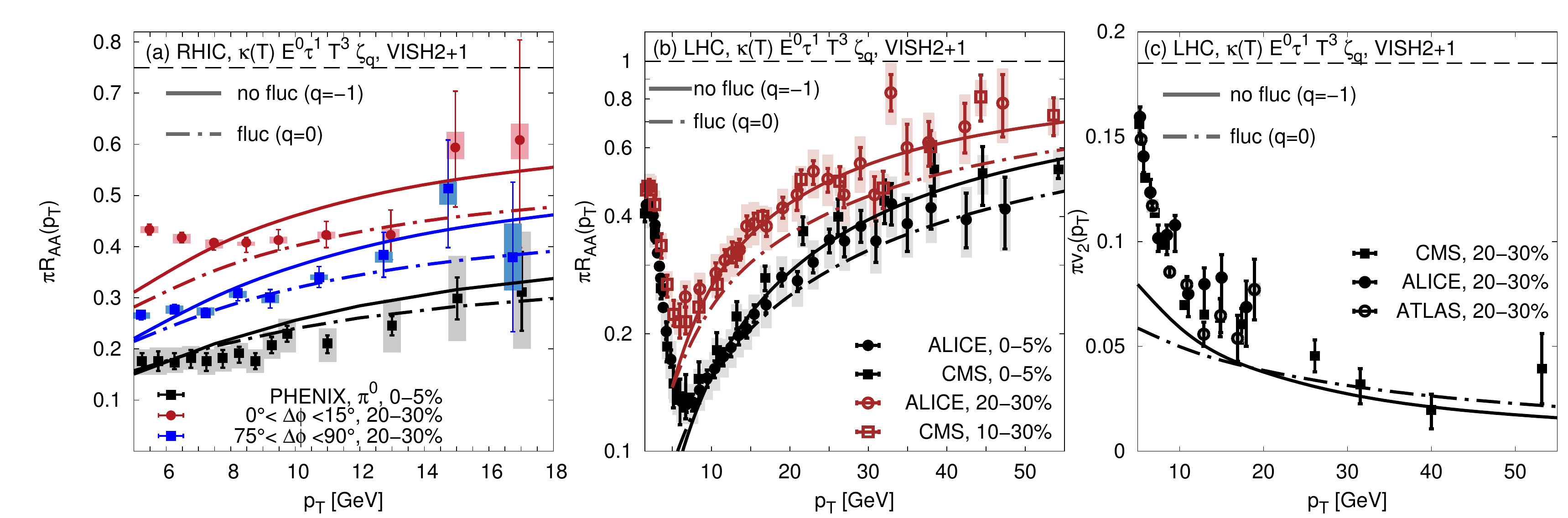}
\caption{Azimuthal jet tomography at RHIC and LHC assuming a pQCD-like
jet-energy loss and a jet-medium coupling $\kappa(T)$ showing an exponential
temperature-dependence as given in Eq.\ (\ref{k_exp}) with ($q=0$) and 
without ($q=-1$) additional jet-energy loss fluctuations considering the bulk 
QGP flow fields from VISH2+1 \protect{\cite{Shen:2011eg}}. Panel (a) shows 
the nuclear modification factor for most central collisions as well as their 
in-and out-of-plane contributions at RHIC, panel (b) depicts the $R_{AA}(p_T)$ 
at LHC, and panel (c) describes the high-$p_T$ elliptic flow at LHC energies. }
\label{Fig12}
\end{center}
\end{figure*}

\begin{figure*}[t]
\begin{center}
\includegraphics[width=6in]{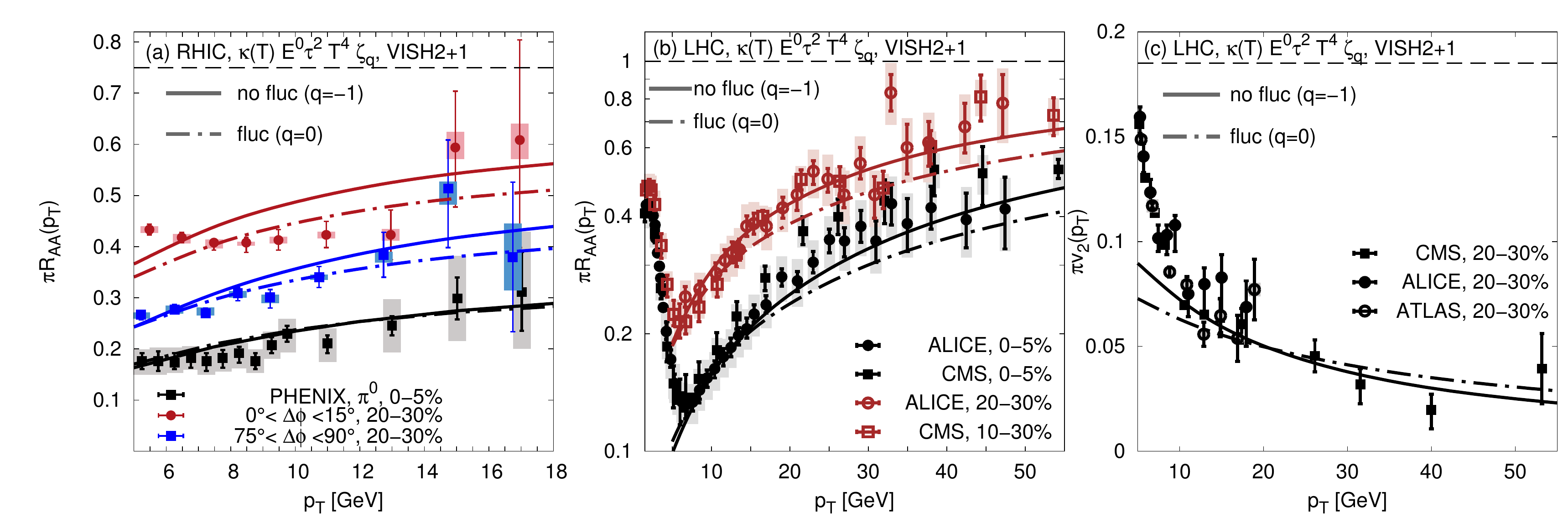}
\caption{Azimuthal jet tomography at RHIC and LHC assuming an AdS-inspired
jet-energy loss scenario with a squared path-length dependence 
and a jet-medium coupling $\kappa(T)$ showing an exponential
temperature-dependence as given in Eq.\ (\ref{k_exp}) with ($q=0$) and 
without ($q=-1$) additional jet-energy loss fluctuations considering the bulk 
QGP flow fields from VISH2+1 \protect{\cite{Shen:2011eg}}. Panel (a) shows 
the nuclear modification factor for most central collisions as well as their 
in-and out-of-plane contributions at RHIC, panel (b) depicts the $R_{AA}(p_T)$ 
at LHC, and panel (c) describes the high-$p_T$ elliptic flow at LHC energies.}
\label{Fig13}
\end{center}
\end{figure*}

We will start the discussion of the LHC results with the pQCD-inspired scenarios 
QCDrad and QCDel in Figs.\ \ref{Fig4} - \ref{Fig8}, while Figs.\ \ref{Fig9} and 
\ref{Fig10} consider the AdS-inspired scenario and the SLTc model.
Figs.\ \ref{Fig11} - \ref{Fig13} then deepen the discussion of a 
temperature-dependent jet-medium coupling $\kappa(T)$.

Fig.\ \ref{Fig4} depicts the central (black) and more peripheral (red) pion nuclear 
modification factors as measured by ALICE (dots) \cite{Abelev:2012hxa} and CMS 
(squares) \cite{CMS:2012aa} in the left panels as well as the high-$p_T$ elliptic 
flow as measured by ALICE (filled dots) \cite{Abelev:2012di}, ATLAS (open dots) 
\cite{ATLAS:2011ah}, and CMS (squares) \cite{Chatrchyan:2012xq} in the right
panels. Those measured data are compared to the results of the QCDrad (upper
panels) and QCDel (lower panels) scenarios for the three different background
fields of viscous ($\eta/s=0.08$) VISH2+1 \cite{Shen:2011eg} (solid), viscous 
($\eta/s=0.08$) RL Hydro \cite{Luzum:2008cw} (dashed-dotted), and the 
$v_\perp=0.6$ blast wave model \cite{GVWH} (dotted) {\em without} jet-energy 
loss fluctuations. The jet-medium coupling constant $\kappa$ is lowered to 
$\sim (40-50)\%$ as compared to RHIC energies to account for running-coupling 
effects \cite{Betz:2012qq,Xu14,WHDG11,Buzzatti:2012dy} explaining the 
``surprising transparency'' \cite{WHDG11} of the LHC QGP (see also Table
\ref{tablesurvey1}). 

A reduction of the effective jet-medium coupling with $\sqrt{s}$ is natural 
\cite{Buzzatti:2012dy,Zakharov:2012fp,Kaczmarek:2004gv} in perturbative QCD 
based jet-energy loss due to vacuum running of both radiative emission and
elastic scattering couplings, 
$\kappa_{QCD} \propto \alpha_s\left(k_\perp^2/[x(1-x)]\right)\alpha_s^2(Q^2)$, 
as a function  of the radiated gluon momentum fraction $x$, the gluon transverse 
momentum $k_\perp$, and the medium momentum transfers $Q$. Lattice QCD 
\cite{Kaczmarek:2004gv} predicts that $\alpha_{\rm effective}(Q,T) = \alpha_{eff}(Q,T)$ 
runs also with the temperature scale.

Fig.\ \ref{Fig4} demonstrates that the QCDrad scenario with a 
$dE/dx=\kappa E^0\tau^1 T^3\zeta_{-1}$ reproduces both the nuclear modification
factors for most central and more peripheral collisions, as well as the high-$p_T$
elliptic flow for either the viscous VISH2+1 or the RL Hydro background fields 
given the uncertainties of the bulk space-time evolution expressed,
amongst others, by the initial conditions, the initial time $\tau_0$, the shear 
viscosity over entropy ratio $\eta/s$, and the the freeze-out time $T_f$. 
Neglecting these uncertainties, as done in Fig.\ \ref{Fig7} below, the high-$p_T$
elliptic flow for VISH2+1 is not well described. Please note that we consider the 
uncertainties in the hydrodynamic prescriptions important and thus conclude that
the QCDrad scenario without jet-energy loss fluctuations provides 
a description of the measured data. The $v_\perp=0.6$ 
blast wave model, however, again fails to describe the measured data. Moreover, 
the QCDrad prescription seems to be favored over the QCDel scenario as for the 
latter one the high-$p_T$ elliptic flow is by $\sim$~2 reduced as compared to 
the QCDrad results.

Fig.\ \ref{Fig5} shows the same comparison as Fig.\ \ref{Fig4}, however, 
depicting one case without jet-energy loss fluctuations (solid) and one
scenario with fluctuations ($q=1$, dashed). Clearly, the jet-energy loss
fluctuations reduce the yield of the nuclear modification factors, both central
and non-central, and the yield of the high-$p_T$ elliptic flow below 
$p_T\le 20$~GeV, while it simultaneously slightly enhances the high-$p_T$ 
elliptic flow above $p_T> 20$~GeV.

There is a certain ambiguity in the yield of both the nuclear modification
factor and the high-$p_T$ elliptic flow that becomes obvious when comparing
Fig.\ \ref{Fig5}(a) to Fig.\ \ref{Fig6}. In Fig.\ \ref{Fig5}(a), we determine
the reduction of the jet-medium coupling by assuming that the 
$R_{AA}(p_T=10\,{\rm GeV})=0.155$, while we supposed in Fig.\ \ref{Fig6} that
$R_{AA}(p_T=10\,{\rm GeV})=0.186$. Both numbers are in line with the current
error bars of the measured date from ALICE and CMS. This comparison 
demonstrates that a larger value for the nuclear modification factor implies 
a lower yield of the elliptic flow.

Neglecting the uncertainties given by the hydrodynamic expansions discussed above 
as we cannot easily assign them a theoretical error bar, Fig.\ \ref{Fig7} reveals 
that both the non-fluctuating and the fluctuating scenarios of QCDrad based on 
the VISH2+1 background fields account for the nuclear modification 
factors on $\chi^2/{\rm d.o.f.}<1.5$ level but fail to describe the high-$p_T$ elliptic flow data 
which is in line with the results of CUJET2.0 \cite{Xu14}. Please note that the 
discrepancies obtained in describing the high-$p_T$ elliptic flow certainly depend 
on the background flow considered, as shown in Fig.\ \ref{Fig4} where the RL Hydro 
scenario is much closer to the measured data. 

In any case, there is a tendency that the high-$p_T$ elliptic flow is too 
small as compared to data. To overcome this ``high-$p_T$ $v_2$ problem'' of 
pQCD-based jet-energy loss prescriptions, Ref.\ \cite{Xu14} suggested that 
in addition to the vacuum running of the perturbative QCD coupling 
$\alpha_s(Q)$, there could well be a running w.r.t.\ the temperature 
$\alpha_{\rm eff}(Q,T)$ \cite{Kaczmarek:2004gv} which could cause modest 
($10-15\%$) variations of the path-averaged coupling in non-central collisions 
with a coupling constant enhanced out-of-plane. To simulate this effect, we include 
an azimuthal dependence of the jet-medium coupling by 
$\kappa(\phi) = \kappa\cdot(1+\vert\sin(\phi)\vert\cdot X)$, 
where $X$ is a value in percentage. 

Fig.\ \ref{Fig8} proves that a small azimuthal variation of $10-15\%$ of the 
jet-medium coupling $\kappa$ is already sufficient to account for the high-$p_T$
$v_2$ problem at LHC energies while simultaneously describing the nuclear modification
factor in- and out-of-plane at RHIC energies as well as the nuclear modification
factors at LHC energies, in line with Ref.\ \cite{Xu14}. However, Fig.\ 
\ref{Fig8} also demonstrates the impact of the bulk QGP background fields. 
While the results for VISH2+1 \cite{Song:2008si,Shen:2010uy} and the RL Hydro 
\cite{Luzum:2008cw} are very similar at RHIC energies, the RL Hydro 
background leads to nuclear modification factors that only touch the lower
bounds of the measured error bars. Please note that we here assume 
$R_{AA}(p_T=10\,{\rm GeV})=0.155$ and jet-energy loss fluctuations. 
Thus, a combined jet-energy loss and bulk QGP background evolution has a 
much larger discriminating power than the two separate prescriptions.

In contrast to the consistent prescription of both RHIC and LHC data for the 
QCDrad scenario considering either no jet-energy loss fluctuations (see Fig.\ \ref{Fig4}) 
or jet-energy loss fluctuations with an additional moderate azimuthal dependence of the 
jet-medium coupling, {\em conformal} AdS-inspired models \cite{Marquet:2009eq,Jia:2011pi} 
based on a $dE/dx\equiv \kappa x^2T^4$ and the SLTc model with an enhanced 
jet-medium coupling around $T_c\sim 170$~GeV fail the extrapolation to LHC
energies for the same backgrounds considered, as shown in Fig.\ \ref{Fig9}.

The reason is that the same jet-medium coupling $\kappa$ is assumed for RHIC 
and LHC in both cases, however for two different reasons: 
In AdS/CFT, $\kappa\propto \sqrt{\lambda}$, where 
$\lambda = 4\pi \alpha_s N_c$ is the 'tHooft coupling that must be 
$\lambda\gg 1$ to ensure applicability of {\em classical} gravity holography.
For {\em conformal} AdS/CFT symmetry, $\lambda$ cannot run. In that 
case, as shown in Fig.\ \ref{Fig9}(a), the AdS prescription over-quenches at
the LHC (which is the well-known ``surprising transparency'' \cite{WHDG11}) and 
leads to a simultaneous enhancement of the high-$p_T$ elliptic flow. This 
over-quenching behaviour was shown even for $\lambda$ as low as $1$ (in static 
backgrounds) and quadratic curvature corrections \cite{Ficnar:2012yu} for
AdS falling string models \cite{Gubser:2008as}. Thus, jet-energy loss 
prescriptions based on {\em conformal} AdS/CFT are ruled out by the rapid
rise of the nuclear modification factor at LHC energies.

The SLTc model \cite{Liao:2008dk}, on the other hand, assumes the dominance of 
jet-energy loss in regions of the QGP with $T\sim T_c\approx 170$~MeV, associating 
the QCD conformal anomaly near $T_c$ with a color magnetic monopole condensation. 
Scattering of color electric charged jets by color magnetic monopoles 
could lead to an enhancement of the jet-energy loss in the QCD crossover 
transition regions that have a higher spatial elliptic eccentricity than 
the average. Following a suggestion of Ref.\ \cite{Liao:2008dk}, 
we simulate this effect by a simple step function of the 
local jet-medium coupling with $\kappa_c = \kappa_c(113<T<173\;{\rm MeV}) = 3\kappa_Q $ 
and $\kappa_Q=\kappa_Q(T\ge173\;{\rm MeV})$. Assuming $\kappa_c/\kappa_Q=3$ 
and a transverse expanding medium, we also observe an over-quenching of the 
nuclear modification factors as shown in Fig.\ \ref{Fig9}(b). Please note that
a generalization of this model with an additional collision energy 
$\sqrt{s}$-dependence weakening the jet-medium coupling at higher $\sqrt{s}$,
and thus considering the running-coupling effect 
\cite{Betz:2012qq,Xu14,WHDG11,Buzzatti:2012dy} with a $\kappa(T,\sqrt{s})$,
will certainly provide an adequate prescription for the $R_{AA}$. To account
for the high-$p_T$ elliptic flow data as well, an additional azimuthal
dependence of the jet-medium coupling, as discussed in Fig.\ \ref{Fig8}, might 
be necessary.

Even though present jet-energy loss models based on AdS/CFT do not consider
fluctuations in the energy loss, we also examine the impact of such additional
jet-energy loss fluctuations for the AdS scenario in Fig.\ \ref{Fig10}(a) and (b).
Again (cf.\ Fig.\ \ref{Fig5}), additional jet-energy loss fluctuations 
cause a stronger jet quenching and a flattening of the high-$p_T$
elliptic flow. 

However, broadening the applicability of holographic models to heavy-ion 
collisions by allowing for {\em non-conformal} AdS/CFT prescriptions that 
enable a running-coupling effect with a reduced jet-medium coupling at LHC 
energies outweigh the over-quenching and result in a simultaneous prescription
of the nuclear modification factors at different centralities as well as
the high-$p_T$ elliptic flow with and without additional jet-energy loss
fluctuations as shown in Fig.\ \ref{Fig10}(c) and (d).
Certainly, such an ansatz requires further generalization of the present holographic 
jet-quenching models to include possibly more general string initial 
conditions and non-conformal geometric deformations \cite{ficnar13,Mia:2012yq}.

Despite the result of Fig.\ \ref{Fig9} showing that the original SLTc model with 
a jet-medium coupling which is non-negligible for large temperatures and 
enhanced in a transition area $\kappa_c(113<T<173\;{\rm MeV})$ is ruled
out by the measured data at the LHC, the fact reported by CUJET2.0 \cite{Xu14} 
and shown in Fig.\ \ref{Fig8} that a moderate azimuthal dependence of the 
jet-medium coupling $\kappa$ with a coupling {\em enhanced out-of-plane}
can result in a simultaneous prescription of the nuclear modification factor 
and the high-$p_T$ elliptic flow at RHIC and LHC supports a jet-medium
coupling enhanced for lower temperatures. The reason is that a jet traversing
out-of-plane will propagate longer through a comparably cooler medium. 

As the original SLTc model does not reproduce the opacity of the LHC medium
appropriately, we consider below an exponentially falling ansatz for the jet-medium
coupling,
\begin{eqnarray}
\hspace*{-3ex}
\kappa(T) = \kappa_1 e^{-b(T-T_1)}\;.
\label{k_exp}
\end{eqnarray}
Here, the coupling is assumed to be zero below a certain temperature $T_1$,
representing the freeze-out, where the coupling peaks at a value of $\kappa_1$
and falls off for larger temperatures to a value of $1/e$ at a temperature $T_e$,
see Fig.\ \ref{Fig11}. 

Figs.\ \ref{Fig12} and \ref{Fig13} depict the results for the nuclear modification
factor and the high-$p_T$ elliptic flow both at RHIC and at LHC energies with 
and without additional jet-energy loss fluctuations, considering the bulk QGP 
flow fields from VISH2+1 \cite{Shen:2011eg} for either a pQCD-based jet-energy 
loss scenario, Fig.\ \ref{Fig12}, or an AdS-inspired prescription with a squared 
path-length dependence, Fig.\ \ref{Fig13}.

Please note that in contrast to previous figures there is only {\em one}
fixing point considered here, $R_{AA}(p_T=7.5\,{\rm GeV},\,{\rm RHIC})=0.2$.
Applying the exponentially falling jet-medium coupling $\kappa(T)$, a reduction
of the {\em effective} jet-medium coupling at the LHC is intrinsic as an LHC-jet
propagates longer through a high-temperature region with a smaller coupling. 
An additional reduction of the jet-medium coupling is not needed. Of course,
the SLTc model also shows such an intrinsic weakening of the jet-medium coupling
at the LHC. However, as proven in Fig.\ \ref{Fig9}, this intrinsic reduction
is not sufficient to account for the transparency at LHC energies 
\cite{Betz:2012qq,Xu14,WHDG11,Buzzatti:2012dy}.

Surprisingly enough, the results shown in Figs.\ \ref{Fig12} and \ref{Fig13} 
with a $T_1=160$~MeV for all scenarios point to having a comparatively low value 
for $T_e$, see Table \ref{tablekappa}, indicating that the high-temperature 
medium is basically transparent \cite{Betz:2012qq,Xu14,WHDG11,Buzzatti:2012dy}. 

Comparing Figs.\ \ref{Fig12} and \ref{Fig13}, both the pQCD-based and the
AdS-inspired models describe the measured data within the present error bars. 
This reinforces our conclusion that the path-length exponent cannot yet be
constrained to a range narrower than $z=[0-2]$. Please note that $z=0$ seems
to be disfavored as shown by the QCDel results but cannot definitely be excluded.

Please also note that the exponential falling $\kappa(T)$ as given by Eq.\ 
(\ref{k_exp}) is {\em one} possible ansatz for a jet-medium coupling describing 
the transparency of the LHC medium (via the $R_{AA}(p_T)$) and the high-$p_T$ 
elliptic flow appropriately. However, other ansaetze might work as well.

\begin{table}[t]
\hspace*{4.2cm}
\begin{tabular}{|c|r|c|c|}
\hline
scenario & $q$ & $T_e$ [MeV]& $\kappa_1(T_1)$\\
\hline
\hline
pQCD-like & -1 & 250 & 1.281\\
\hline
pQCD-like & 0 & 270 & 2.134\\
\hline
AdS-inspired & -1 & 220 & 0.589\\
\hline
AdS-inspired & 0 & 230 & 0.956\\
\hline
\end{tabular}
\caption{The parameters for the exponential temperature-dependent
jet-medium coupling $\kappa(T)$ as given by Eq.\ (\ref{k_exp}) and shown
in Figs.\ \ref{Fig12} and \ref{Fig13} with $T_1=160$ MeV and 
$\kappa_e(T_e)=1/e=0.37$ [in units of $1/(\hbar c)^{(1+z)}$].}
\label{tablekappa}
\end{table}

Finally we note that at both RHIC and LHC energies, the magnitude of the nuclear
modification factors in the intermediate (``IM'') $2<p_T<7$~GeV kinematic region 
is under-predicted by all jet-quenching models considered here. This ``IM'' 
region interpolates between the perfect fluid low-$p_T<2$~GeV infrared (``IR'') 
range and the high-$p_T>7$~GeV ultraviolet (``UV'') perturbative QCD jet-quenching 
range. A proper theory of jet quenching in the non-equilibrium QGP ``IM'' range 
remains a formidable challenge.

\section{Conclusions} 
We compare recent data on the nuclear modification factor and the
high-$p_T$ elliptic flow measured at RHIC \cite{Adare:2012wg} and LHC
energies
\cite{Abelev:2012hxa,Abelev:2012di,CMS:2012aa,Chatrchyan:2012xq,ATLAS:2011ah}
to a broad class of jet-energy independent energy-loss models (see
Table \ref{tablesurvey1}) with $dE/dx= \kappa(T) E^{a=0} x^z T^c \zeta_q$, 
labelled by $(z,c,q)$, including jet-energy loss fluctuations for $q>-1$. In
particular, we study (a) a linear, radiative pQCD-like jet-energy loss
with running coupling \cite{Betz:2012qq,Xu14,Buzzatti:2012dy}, (b) a
linear, elastic pQCD-like jet-energy loss with running coupling
\cite{WHDG11}, (c) an AdS/CFT-inspired, quadratic jet-energy loss, (d)
a $T_c$-dominated energy-loss model (SLTc) \cite{Liao:2008dk}, (e) an
energy-loss prescription based on a moderate azimuthal dependence of
the jet-medium coupling \cite{Xu14}, and (f) an energy-loss scenario
with a temperature-dependent jet-medium coupling $\kappa(T)$ dropping
exponentially for large temperatures.  All those models are combined
with several recent transverse and Bjorken expanding, collective flow
backgrounds \cite{Song:2008si, Shen:2010uy, Shen:2011eg, Luzum:2008cw, GVWH}.

We find (see Table \ref{tablesurvey2}) that (1) running coupling energy-loss 
models with $(0,1,3)$ motivated by perturbative QCD appear to be favored, 
(2) {\em conformal} AdS/CFT-inspired jet-energy loss scenarios are ruled out by 
the reduction of coupling required to fit LHC data, and (3) a realistic [$(2+1)$d] 
QGP flow background is essential to account for the dependence of the data on 
transverse momentum $p_T$, azimuth $\phi$, impact parameter $b$, and collision 
energy $\sqrt{s}$. 

We further explore different possible deformations of 
the models that could reduce the discrepancies with the combined RHIC and LHC 
data. The simplest solution utilizes either (viscous) VISH2+1 
\cite{Song:2008si,Shen:2011eg} or RL \cite{Luzum:2008cw} hydrodynamic
fields and corresponds to a radiative pQCD-like energy loss with
$dE/dx= \kappa x^1 T^3 \zeta_{-1}$, $(1,3,-1)$, with $\zeta_{-1}=1$
neglecting jet-energy loss fluctuations.  While this fits, it is theoretically 
not compelling. A second solution includes those energy-loss fluctuations ($q=0,1$) 
for $(1,3,q)$ pQCD-like modelling, but allows for $\sim 10-15$~\% variation of 
the jet-medium coupling along different paths relative to the reaction plane. 
The exact value of the variation depends on the QGP flow fields used
\cite{Song:2008si,Shen:2011eg,Luzum:2008cw}. This class of solutions is similar 
to the one recently proposed by the more detailed CUJET2.0 model \cite{Xu14}. It 
remains to be seen whether the combined temperature and scale running of jet-medium 
coupling $\alpha_{\rm eff}(Q,T)$ \cite{Xu14,Kaczmarek:2004gv} can be more 
rigorously justified.

A third class of $(1,3,q)$ solutions assumes a more radical
temperature-dependent jet-medium coupling $\kappa(T)$ with not only an
enhancement of the coupling near $T_c$, in the spirit of the SLTc
scenario, but requiring an exponential suppression of $\kappa(T)$
at high temperatures that is rather puzzling from both a pQCD and an
AdS points of view. A fourth class of solutions also assumes a 
temperature-dependent jet-medium coupling $\kappa(T)$ with an enhancement 
of the coupling near $T_c$ as given by the SLTc scenario, but requiring an 
additional reduction of the magnitude of $\kappa(T)$ at the LHC.

Finally, a fifth class of AdS-inspired solutions with a quadratic jet-path length 
dependence $dE/dx=\kappa(T) x^2 T^4 \zeta_q$ premises a strong {\em non-conformal} 
reduction of jet-medium coupling by a factor of two at LHC energies. Thus far, 
no holographic model has predicted such strong conformal breaking effects. 

Please see Table \ref{tablesurvey1} for a summary of the 
relative success and failure of the different models surveyed in 
Table \ref{tablesurvey2}.

Given the current landscape of jet-medium modelling in Table \ref{tablesurvey2} and
the uncertainties in justifying deformations of current models required to fit 
the data, especially the LHC high-$p_T$ elliptic moment $v_2(p_T)$, we cannot 
constrain the path-length exponent $z$ of the jet-energy loss to a 
range narrower than $z=[0-2]$.

\section{Acknowledgments} 
We are especially grateful to P.\ Romatschke, U.\ Heinz, and C.\ Shen for making 
their hydrodynamic field grids available. Discussions with  J. Xu, A.\ Ficnar, 
A.\ Buzzatti, W.\ Horowitz, J.\ Liao, D.\ Molnar, and X.-N.\ Wang in the JET 
Collaboration have been particularly valuable. BB acknowledges financial support 
received from the Helmholtz International Centre for FAIR within the framework
of the LOEWE program (Landesoffensive zur Entwicklung Wissenschaftlich-\"Okonomischer 
Exzellenz) launched by the State of Hesse. MG acknowledges support from the 
US-DOE Nuclear Science Grant No.\ DE-FG02-93ER40764 and No.\ DE-AC02-05CH11231 
within the framework of the JET Topical Collaboration \cite{JET}. The authors 
also thank the Yukawa Institute for Theoretical Physics, Kyoto University, where 
part of this work was completed during the YITP-T-13-05 on "New Frontiers in QCD". 
MG is grateful for partial support from the MTA Wigner RCP, Budapest, during the 
second half of his sabbatical leave in 2014, where this work was finalized.

\newpage

\section{Appendix}
\subsection*{Jet+bulk models used in the present survey}
\begin{table}[h]
\begin{tabular}{|c|c|c|c|c|c|c|c|}
\hline
$\#$ & name & fluct.\ & $(z,c,q)$ & temp.\ profile& $\kappa_{\rm RHIC}$ & $\kappa_{\rm LHC}$ & Fig.\ $\#$\\
\hline
1 & QCDrad & no & $(1,3,-1)$ & VISH2+1 & 0.380 & 0.167 & 1,4,5\\
\hline
1a & QCDrad & no & $(1,3,-1)$ & VISH2+1 & 0.380 & 0.136 & 1,6\\
\hline
2 & QCDrad & no & $(1,3,-1)$ & RL Hydro & 0.477 & 0.241 & 1,4\\
\hline
3 & QCDrad & no & $(1,3,-1)$ & $v=0.6$ & 3.182 & 2.096 & 1,4\\
\hline
4 & QCDel & no & $(0,2,-1)$ & VISH2+1 & 0.887 & 0.483 & 1,4\\
\hline
5 & QCDel & no & $(0,2,-1)$ & RL Hydro & 1.497 & 0.906 & 1,4\\
\hline
6 & QCDel & no & $(0,2,-1)$ & $v=0.6$ & 5.713 & 5.024 & 1,4\\
\hline
7 & AdS & no & $(2,4,-1)$ & VISH2+1 & 0.092 & 0.092 & 1,9\\
\hline
8 & AdS & no & $(2,4,-1)$ & RL Hydro & 0.145 & 0.145 & 1,9\\
\hline
9 & AdS & no & $(2,4,-1)$ & $v=0.6$ & 1.911 & 1.911 & 1,9\\
\hline
10 & SLTc & no & $(1,3,-1)$ & VISH2+1 & 0.167 & 0.167 & 1,9\\
\hline
11 & SLTc & no & $(1,3,-1)$ & RL Hydro & 0.330 & 0.330 & 1,9\\
\hline
12 & SLTc & no & $(1,3,-1)$ & $v=0.6$ & 1.591 & 1.591 & 1,9\\
\hline
13 & QCDrad & yes & $(1,3,+1)$ & VISH2+1 & 0.718 & 0.349 & 2,5\\
\hline
13a & QCDrad & yes & $(1,3,+1)$ & VISH2+1 & 0.718 & 0.269 & 2,6\\
\hline
14 & QCDel & yes & $(1,3,+1)$ & VISH2+1 & 1.615 & 1.024 & 2,5\\
\hline
15 & AdS & yes & $(2,4,+1)$ & VISH2+1 & 0.283 & 0.283 & 2,10(a,b)\\
\hline
16 & ncAdS & no & $(2,4,-1)$ & VISH2+1 & 0.092 & 0.047 & 2,10(c,d)\\
\hline
17 & ncAdS & yes & $(2,4,+1)$ & VISH2+1 & 0.283 & 0.111 & 2,10(c,d)\\
\hline
18 & $\kappa(\phi)$ QCDrad & yes & $(1,3,0)$ & VISH2+1 & 0.543 & 0.235 & 8\\
\hline
19 & $\kappa(\phi)$ QCDrad & yes & $(1,3,0)$ & RL Hydro & 0.776 & 0.345 & 8\\
\hline
20 & exp.\ $\kappa(T)$ QCDrad & no & $(1,3,-1)$ & VISH2+1 & $\kappa_1$=1.281 & $\kappa_1$=1.281 & 12\\
\hline
21 & exp.\ $\kappa(T)$ QCDrad & yes & $(1,3,0)$ & VISH2+1 & $\kappa_1$=2.134 & $\kappa_1$=2.134 & 12\\
\hline
22 & exp.\ $\kappa(T)$ ncAdS & no & $(2,4,-1)$ & VISH2+1 & $\kappa_1$=0.589 & $\kappa_1$=0.589 & 13\\
\hline
23 & exp.\ $\kappa(T)$ ncAdS & yes & $(2,4,0)$ & VISH2+1 & $\kappa_1$=0.956 & $\kappa_1$=0.956 & 13\\
\hline
\end{tabular}
\caption{Parameters of the jet-energy loss models and bulk-temperature fields 
considered in the present survey. The columns show the model identifier name, 
whether jet-energy loss fluctuations are considered, the path-length exponent $z$, 
the temperature exponent $c$, and the energy-loss fluctuation parameter $q$ 
as well as the bulk-temperature $T(\vec{x},t)$ field assumed. The effective jet-medium 
coupling at RHIC obtained by a single fit to central $R_{AA}(p_T=7.5\,{\rm GeV})$ 
data are listed in column 6. Column 7 shows the values of the LHC jet-medium 
coupling used in the figures listed in column 8. See text for details.}
\label{tablesurvey1}
\end{table}

\newpage

\begin{table}[h]
\begin{tabular}{|c|c|c|c|c|c|c|c|}
\hline
Scenario & \multicolumn{3}{c|}{RHIC} & \multicolumn{3}{c|}{LHC}  & \multicolumn{1}{c|}{Score}\\
$\#$ & $R_{\rm AA}^{\rm centr}$ & $R_{\rm AA}^{\rm in, periph}$ & $R_{\rm AA}^{\rm out, periph}$ &
$R_{AA}^{\rm centr}$ & $R_{AA}^{\rm periph}$ & $v_2^{\rm periph}$ & Sum \\
\hline
1 & $\checkmark$ & $\checkmark$ & $\checkmark$ & $\checkmark$ & $\checkmark$ & $(\checkmark)$ & 5\\  
\hline
1a & $\checkmark$ & $\checkmark$ & $\checkmark$ & $(\checkmark)$ & $(\checkmark)$ & $(\checkmark)$ & 3\\ 
\hline 
2 & $\checkmark$ & $\checkmark$ & $\checkmark$ & $\checkmark$ & $\checkmark$ & $(\checkmark)$ & 5\\  
\hline
3 & $(\checkmark)$ & $\checkmark$ & no & $\checkmark$ & $\checkmark$ & no & 1\\ 
\hline 
4 & $\checkmark$ & $\checkmark$ & $\checkmark$ & $(\checkmark)$ & $(\checkmark)$ & $(\checkmark)$ & 3 \\  
\hline
5 & $\checkmark$ & $\checkmark$ & $\checkmark$ & $\checkmark$ & $(\checkmark)$ & $(\checkmark)$ & 4\\  
\hline
6 & $\checkmark$ & no & $\checkmark$ & $(\checkmark)$ & $(\checkmark)$ & no & 0\\  
\hline
7 & $\checkmark$ & $\checkmark$ & $\checkmark$ & no & no & $\checkmark$ & 2\\  
\hline
8 & $\checkmark$ & $\checkmark$ & $\checkmark$ & no & no & no & 0\\ 
\hline 
9 & $\checkmark$ & $\checkmark$ & no & no & no & $(\checkmark)$ & -1 \\  
\hline
10 & $\checkmark$ & $\checkmark$ & $\checkmark$ & no & no & $\checkmark$ &2\\  
\hline
11 & $\checkmark$ & $\checkmark$ & $\checkmark$ & no & no & $\checkmark$ & 2 \\  
\hline
12 & $(\checkmark)$ & no & no & no & no & no & -5\\  
\hline
13 & $\checkmark$ & $(\checkmark)$ & $(\checkmark)$ & $(\checkmark)$ & no & $(\checkmark)$ & 0 \\  
\hline
13a & $\checkmark$ & $(\checkmark)$ & $(\checkmark)$ & $\checkmark$ & $(\checkmark)$ & $(\checkmark)$ & 2 \\  
\hline
14 & $\checkmark$ & no & no & $\checkmark$ & no & no & -2 \\  
\hline
15 & $\checkmark$ & $\checkmark$ & $(\checkmark)$ & no & no & $(\checkmark)$ & 0\\  
\hline
16 & $\checkmark$ & $(\checkmark)$ & $\checkmark$ & $\checkmark$ & $\checkmark$ & $\checkmark$ & 5 \\  
\hline
17 & $\checkmark$ & $\checkmark$ & $(\checkmark)$ & no & no & $\checkmark$ & 1\\  
\hline
18 & $\checkmark$ & $\checkmark$ & $\checkmark$ & $\checkmark$ & $\checkmark$ & $\checkmark$ & 6 \\  
\hline
19 & $\checkmark$ & $\checkmark$ & $\checkmark$ & no & no & $(\checkmark)$ & 1\\  
\hline
20 & $\checkmark$ & $(\checkmark)$ & $\checkmark$ & $\checkmark$ & $\checkmark$ & $\checkmark$ & 5\\  
\hline
21 & $\checkmark$ & $\checkmark$ & $(\checkmark)$ & $(\checkmark)$ & no & $\checkmark$ & 1\\  
\hline
22 & $\checkmark$ & $\checkmark$ & $(\checkmark)$ & $\checkmark$ & $\checkmark$ & $\checkmark$ & 5\\  
\hline
23 & $\checkmark$ & $\checkmark$ & $\checkmark$ & $(\checkmark)$ & no & $\checkmark$ & 3 \\ 
\hline
\end{tabular}
\caption{Relative matrix of success and failure of the jet+bulk models 
surveyed in Table \ref{tablesurvey1} based on comparisons of the results to
RHIC and LHC data for $R_{AA}(p_T, \phi, b, \sqrt{s})$ and high-$p_T$ 
$v_{2}(p_T, \phi, b, \sqrt{s})$. The last column shows a score given by
Score = number of checks - number of no's. A "(check)" indicates an inconclusive 
judgement of success and is given a zero weight. Model 18 has highest 
score 6, while the models 1, 3, 16, 20, and 22 are tied at score 5. See text for 
discussion.}
\label{tablesurvey2}
\end{table}

\end{document}